\documentclass[twocolumn]{aastex63}

\DeclareFontFamily{U}{dutchcal}{\skewchar\font=45 }
\DeclareFontShape{U}{dutchcal}{m}{n}{<-> s*[1.0] dutchcal-r}{}
\DeclareFontShape{U}{dutchcal}{b}{n}{<-> s*[1.0] dutchcal-b}{}
\DeclareMathAlphabet{\mathlcal}{U}{dutchcal}{m}{n}
\SetMathAlphabet{\mathlcal}{bold}{U}{dutchcal}{b}{n}

\usepackage{amssymb,amsmath,verbatim,mathtools,needspace,enumitem,graphicx,physics,microtype,mathrsfs}
\usepackage{amssymb}
\def\gtsima{$\; \buildrel > \over \sim \;$}
\def\ltsima{$\; \buildrel < \over \sim \;$}
\def\gtrsim{\lower.5ex\hbox{\gtsima}}

\def\lesssim{\lower.5ex\hbox{\ltsima}}

\usepackage{xcolor}

\graphicspath{{./}{./}}

\submitjournal{ApJ}

\begin{document}

\title{\large Constraining the fraction of binary black holes formed in isolation and young star clusters with gravitational-wave data}

\author{Yann Bouffanais}
\email{yann.bouffanais@pd.infn.it}
\affiliation{Dipartimento di Fisica e Astronomia `G. Galilei', University of Padova, Vicolo dell'Osservatorio 3, I--35122, Padova, Italy}
\affiliation{INFN, Sezione di Padova, Via Marzolo 8, I--35131, Padova, Italy}

\author[0000-0001-8799-2548]{Michela Mapelli}
\email{michela.mapelli@unipd.it}
\affiliation{Dipartimento di Fisica e Astronomia `G. Galilei', University of Padova, Vicolo dell'Osservatorio 3, I--35122, Padova, Italy}
\affiliation{INFN, Sezione di Padova, Via Marzolo 8, I--35131, Padova, Italy}
\affiliation{INAF--Osservatorio Astronomico di Padova, Vicolo dell'Osservatorio 5, I--35122, Padova, Italy}
\affiliation{Institut f\"ur  Astro- und Teilchenphysik, Universit\"at Innsbruck, Technikerstrasse 25/8, A--6020, Innsbruck, Austria}

\author[0000-0002-0933-3579]{Davide Gerosa}
\affiliation{School of Physics and Astronomy and Institute for Gravitational Wave Astronomy, University of Birmingham, \\Birmingham, B15 2TT, UK}

\author{Ugo N. Di Carlo}
\affiliation{Dipartimento di Fisica e Astronomia `G. Galilei', University of Padova, Vicolo dell'Osservatorio 3, I--35122, Padova, Italy}
\affiliation{INFN, Sezione di Padova, Via Marzolo 8, I--35131, Padova, Italy}
\affiliation{Dipartimento di Scienza e Alta Tecnologia, University of Insubria, Via Valleggio 11, I--22100, Como, Italy}

\author{Nicola Giacobbo}
\affiliation{Dipartimento di Fisica e Astronomia `G. Galilei', University of Padova, Vicolo dell'Osservatorio 3, I--35122, Padova, Italy}
\affiliation{INFN, Sezione di Padova, Via Marzolo 8, I--35131, Padova, Italy}
\affiliation{INAF--Osservatorio Astronomico di Padova, Vicolo dell'Osservatorio 5, I--35122, Padova, Italy}

\author{Emanuele Berti}
\affiliation{Department of Physics and Astronomy, Johns Hopkins University, 3400 N. Charles Street, Baltimore, MD 21218, US}

\author{Vishal Baibhav}
\affiliation{Department of Physics and Astronomy, Johns Hopkins University, 3400 N. Charles Street, Baltimore, MD 21218, US}

\begin{abstract}
Ten binary black-hole mergers have already been detected during the first two observing runs of advanced LIGO and Virgo, and many more are expected to be observed in the near future. This opens the possibility for gravitational-wave astronomy to better constrain the properties of black hole binaries, not only as single sources, but as a whole astrophysical population. In this paper, we address the problem of using gravitational-wave measurements to estimate the proportion of merging black holes produced either via isolated binaries or binaries evolving in young star clusters. To this end, we use a Bayesian hierarchical modeling approach applied to catalogs of merging binary black holes generated using state-of-the-art population synthesis and N-body codes. In particular, we show that, although current advanced LIGO/Virgo observations only mildly constrain the mixing fraction $\mathlcal{f} \in [0,1]$ between the two formation channels, we expect to narrow down the fractional errors on $\mathlcal{f}$ to $10-20\%$ after a few hundreds of detections.
\end{abstract}


\section{Introduction}
\label{sec_1}
The first two observing runs of the LIGO/Virgo collaboration (LVC) led to the detection of ten binary black holes (BBHs, \citealt{abbottO1,abbottO2}) and one binary neutron star (BNS, \citealt{abbottGW170817,abbottmultimessenger}). The third observing run will significantly boost this sample: several tens of new BBH detections and few BNS detections are expected in the coming months. The growing sample of merging BBHs is expected to provide key information on their mass, spin and local merger rate \citep{abbottO2popandrate}.

One of the main open questions about BBHs concerns their formation channel(s). Several possible scenarios have been proposed in the last decades. The {\emph{isolated evolution}} of a massive binary star can lead to the formation of a merging BBH through a common envelope episode \citep{bethe1998,belczynski2002,belczynski2014,belczynski2016,dominik2013, mennekens2014, spera2015, eldridge2016,eldridge2017, mapelli2017, mapelli2018a, stevenson2017, giacobbo2018b, kruckow2018,spera2019,mapelli2019,eldridge2019} or via chemically homogeneous evolution \citep{marchant2016,mandel2016}.

Alternatively, several {\emph{dynamical processes}} can trigger the formation of a BBH and influence its subsequent evolution to the final merger (see \citealt{mapelli2018b} for a recent review on the subject). For example, the Kozai-Lidov dynamical mechanism \citep{kozai1962,lidov1962} might significantly affect the formation of eccentric BBHs in triple stellar systems (e.g. \citealt{antonini2012,antonini2016,kimpson2016,antonini2017}). Similarly, dynamical exchanges and three- or multi-body scatterings  are expected to lead to the formation and dynamical hardening of BBHs in dense stellar systems, such as globular clusters \citep{portegieszwart2000, oleary2006, sadowski2008, downing2010, downing2011, rodriguez2015, rodriguez2016a, rodriguez2016b, rodriguez2018, askar2017, samsing2018a,samsing2018b, fragione2018}, nuclear star clusters \citep{oleary2009, antonini2012, antonini2016, petrovich2017,stone2017a,stone2017b,rasskazov2019} and young star clusters \citep{banerjee2010, mapelli2013,ziosi2014, mapelli2016, banerjee2017, banerjee2018, dicarlo2019, kumamoto2019}. Other formation mechanisms include black hole (BH) pairing in extreme gaseous environments  (like AGN disks, e.g. \citealt{mckernan2012,mckernan2014,mckernan2018,bartos2017,tagawa2018}). Finally, primordial BHs of non-stellar origin may form binaries through dynamical processes \citep{carr1974,carr2016,sasaki2016,bird2016,inomata2017,inayoshi2016,scelfo2018}.

Each formation channel leaves its specific imprint on the properties of BBHs. In particular, dynamically formed BBHs are expected to have larger masses than isolated BBHs (e.g. \citealt{dicarlo2019}), because dynamical exchanges favour the formation of more massive binaries \citep{hills1980}. Several evolutionary processes in isolated binary systems (tides, mass transfer) tend to align the individual spins with the orbital angular momentum of the binary, while only supernova kicks can tilt the spins significantly in isolated binaries \citep{kalogera2000,gerosa2013,gerosa2018,oshaughnessy2017}. In contrast, dynamical exchanges are expected to reset any memory of previous alignments; thus, dynamically formed BBHs are expected to have isotropically oriented spins. Finally, dynamically formed BBHs (especially Kozai-Lidov triggered systems) might develop larger eccentricities than isolated BBHs. Eccentricities are larger and easier to measure at the low frequencies accessible to space-based interferometers such as LISA \citep{nishizawa2016_1, breivik2016, nishizawa2016_2}, but in some cases they may be significant even in the advanced LIGO (aLIGO) and advanced Virgo (aVirgo) band (e.g. \citealt{antonini2017,zevin2019}).

Thus, BH masses, spins and eccentricities are key features to differentiate between binary formation channels. To achieve this goal, BBH populations predicted by models should be contrasted with gravitational-wave (GW) data, by means of a suitable model-selection framework. Several methodological approaches can be found in the literature \citep{stevenson2015,stevenson2017b,gerosa2017,vitale2017,zevin2017,talbot2017,talbot2018,taylor2018,abbottO2popandrate,fishbach2017,fishbach2018,wysocki2018, roulet2019,kimball2019}.  %
For example, \cite{stevenson2017b} use of a hierarchical analysis in order to combine multiple GW observations of BBH spin--orbit misalignments, to give constraints on the fractions of BBHs forming through different channels. Similarly, \cite{zevin2017} apply a hierarchical Bayesian model to mass measurements from mock GW observations. They compare populations obtained with isolated binary evolution and with Monte Carlo simulations of globular clusters and show that they can distinguish between the two channels with $\mathcal{O}(100)$ GW observations. \cite{taylor2018} use banks of compact-binary population synthesis simulations to train a Gaussian-process emulator that acts as a prior on observed parameter distributions (e.g. chirp mass, redshift, rate). Based on the results of the emulator, a hierarchical population inference framework allows to extract information on the underlying astrophysical population. Alternative approaches consist in model-independent inference based on clustering of source parameters (e.g. \citealt{mandel2015,mandel2017,powell2019}).

Here, we follow a standard Bayesian model-selection approach (cf. e.g. \citealt{sesana2010,gair2010,gerosa2017}), properly including selection effects \citep{mandel2018} and posterior distributions, exploiting both full aLIGO/aVirgo data and mock samples for future forecasts.
As for the astrophysical models, we compare BBHs from isolated binary evolution with dynamically formed BBHs. For the first time, we apply model selection to dynamically formed BBHs in young star clusters \citep{dicarlo2019}. Young star clusters are intriguing dynamical environments for BBHs, because massive stars (which are BH progenitors) form preferentially in young star clusters in the nearby Universe \citep{lada2003,portegieszwart2010}. On the other hand, simulating BBHs in young star clusters has a high computational cost, because it requires direct N-body simulations combined with binary population synthesis. Both isolated binaries and dynamically formed ones are evolved through the {\sc mobse} population-synthesis code \citep{giacobbo2018a}, which includes state-of-the-art modelling of stellar winds, supernova prescriptions and binary evolution.

\section{Distributions of astrophysical sources}
\label{sec_2}

\subsection{Isolated formation channel of BBHs}
\label{Sec_cat_iso}
  We simulate isolated BBHs using the binary population-synthesis code  \textsc{mobse} \citep{mapelli2017,giacobbo2018a}.
 \textsc{mobse} includes single stellar evolution through polynomial fitting formulas as described in \cite{hurley2000} and binary evolution processes (mass transfer, tidal evolution, common envelope, GW decay, etc.) as described in \cite{hurley2002}.  The main differences between \textsc{mobse} and \textsc{bse} are the following (cf. \citealt{giacobbo2018b} for additional details).

Mass loss by stellar winds of massive hot stars (O- and B-type stars, luminous blue variables and Wolf-Rayet stars) is implemented in \textsc{mobse} as $\dot{M}\propto{}Z^{\eta{}}$ \cite[and references therein]{chen2015}, where $Z$ is the stellar metallicity and
\begin{equation}\label{eq:gamma}
  \eta{}=\begin{cases}
  0.85, & \textrm{if}~\Gamma_e<2/3\\
  2.45-2.4\,{}\Gamma{}_e & \textrm{if}~2/3\leq{}\Gamma_e{}\leq{}1\\
  0.05\,{} & \textrm{if}~\Gamma_e>1, 
  \end{cases}
  \end{equation}
where $\Gamma_e=L_\ast{}/L_{\rm Edd}$ is the Eddington factor, $L_\ast{}$ is the current stellar luminosity, and $L_{\rm Edd}$ is the Eddington luminosity \citep{graefener2011}.
The mass of a compact object depends on the final mass and core mass of the progenitor star through fitting formulas which describe the outcome of electron-capture supernovae (see \citealt{giacobbo2019}), core-collapse supernovae (see \citealt{fryer2012}) and pair-instability or pulsational pair-instability supernovae (see \citealt{spera2017}). In this paper, we adopt the delayed model for core collapse supernovae (see \citealt{fryer2012}).
These prescriptions enable us to obtain a BH mass distribution which is consistent with GW data\footnote{Prescriptions that do not account for the dependence of BH mass on progenitor’s metallicity and models that do not include pair instability and pulsational pair instability supernovae are in tension with data. The former because they cannot explain the formation of BHs with mass $>30$ M$_\odot$, the latter because they predict too many BHs with mass $>50$ M$_\odot$.} from the first and second observational runs of aLIGO and aVirgo \citep{abbottO2,abbottO2popandrate}.

The natal kick of a neutron star is drawn from a Maxwellian distribution with 1-dimensional root-mean square $\sigma{}=15$ and $265$ km s$^{-1}$ for an electron-capture and a core-collapse supernova, respectively (see \citealt{hobbs2005} and \citealt{giacobbo2019} for more details). The natal kick of a BH is calculated as $v_{\rm BH}=v_{\rm NS}\,{}(1-f_{\rm fb})$, where $v_{\rm NS}$ is a random number extracted from the same Maxwellian distribution as neutron stars born from core-collapse supernovae, while $f_{\rm fb}$ is the fraction of mass that falls back to a BH, estimated as in \cite{fryer2012}. 

In this paper, we consider a sample of $10^7$ binaries simulated with \textsc{mobse} with metallicity $Z=0.002\simeq Z_\odot/10$ %
(the effect of varying the metallicity will be tackled in a forthcoming publication).

The primary mass is randomly drawn from a \cite{kroupa2001} initial mass function between $m_1=5$ M$_\odot$ and 150 M$_\odot$, while the secondary is randomly drawn from the mass ratio $q$
\begin{equation}
	\mathfrak{F}(q)~ \propto ~q^{-0.1} \qquad ~~~\mathrm{with}~~~q = \frac{m_2}{m_1}~ \in [0.1-1]~.
\end{equation}
As suggested by observations \citep{sana2012}, initial orbital periods $P$ and eccentricities $e$ are randomly drawn  from 

\begin{align}
	&\mathfrak{F}(\mathscr{P}) ~\propto~ (\mathscr{P})^{-0.55} ~~\mathrm{with}~ \mathscr{P} = \mathrm{log_{10}}(P/\mathrm{day}) \in [0.15-5.5]\,,
\\
	&\mathfrak{F}(e) ~\propto ~e^{-0.42} \qquad ~~\mathrm{with}~~~ 0\leq e < 1.
\end{align}

For this paper we adopted common-envelope ejection efficiency $\alpha{}=3$ , while the envelope concentration $\lambda{}$ is derived by \textsc{mobse} as described by \cite{claeys2014}.

From these population-synthesis simulations we obtain 31879 BBH mergers (hereafter referred to as ``isolated BBHs'') which merge within a Hubble time $t_{\rm H}=14$ Gyr.

\subsection{Dynamical formation channel of BBHs in young star clusters}
\label{Sec_cat_dyn}

The dynamically formed BBHs were obtained by means of direct N-body simulations with \textsc{nbody6++gpu} \citep{wang2015} %
coupled to   \textsc{mobse}  \citep{dicarlo2019}.
We have, therefore, the very same population-synthesis recipes in both the isolated binary simulations and the dynamical simulations.

The initial conditions were obtained with \textsc{McLuster} \citep{kuepper2011}.%
 The distributions of dynamical BBHs discussed in this paper are obtained from 4000 simulations of young star clusters with fractal initial conditions. We chose to simulate young star clusters because most stars (and especially massive stars) are thought to form copiously in these environments (e.g. \citealt{lada2003,portegieszwart2010}). The assumption of fractal initial conditions mimics the clumpiness and asymmetry of observed star forming regions (e.g. \citealt{gutermuth2005}). Each star cluster's mass was randomly drawn from a distribution $dN/dM_{\rm SC}\propto{}M_{\rm SC}^{-2}$, consistent with the observed mass function of young star clusters in the Milky Way \citep{lada2003}. Thus, our simulated star clusters represent a synthetic young star-cluster population of Milky Way-like galaxies.

The initial binary fraction in each star cluster is $f_{\rm bin}=0.4$. While observed young star clusters can have larger values of $f_{\rm bin}$ (up to $\sim{}0.7$, \citealt{sana2012}), $f_{\rm bin}=0.4$ is close to the maximum value considered in state-of-the-art simulations, as $f_{\rm bin}$ is the bottleneck of direct N-body simulations. 
Initial stellar and binary masses, orbital periods and orbital eccentricities are generated as described in Sec.~\ref{Sec_cat_iso}, to guarantee a fair comparison. For the same reason, all the simulated star clusters have stellar metallicity $Z=0.002$, the same as isolated binaries.

Each star cluster feels the tidal field of a Milky Way-like galaxy and is assumed to be on a circular orbit with radius similar to the Sun's orbital radius.
Star clusters are simulated for $\sim{}100$ Myr, corresponding to a conservative assumption for the lifetime of a young star cluster. We refer to \cite{dicarlo2019} for a more detailed discussion of our dynamical models and assumptions.

From these dynamical simulations we obtain 229 BBHs (hereafter dynamical BBHs) which merge within a Hubble time $t_{\rm H}=14$ Gyr. We stress that dynamical simulations are computationally more expensive than population-synthesis runs and our sample of merging BBHs is one of the largest ever obtained from direct N-body simulations with realistic binary evolution.

Dynamical BBHs belong to two families. About 47\% of all merging BBHs in the simulated star clusters come from original binaries (hereafter, original BBHs), i.e. they form from the evolution of stellar binaries which were already present in the initial conditions. Such original binaries evolve in a star cluster, thus they are affected by close-by encounters with other stars (which can change their semi-major axis and eccentricity), but otherwise behave similarly to BBHs formed in isolation.

The remaining 53\% of all merging BBHs in the simulated star clusters form via dynamical exchanges (hereafter, exchanged BBHs). Dynamical exchanges are three-body encounters between a binary system and a single object, during which the single object exchanges with one of the members of the binary system. BHs are tremendously efficient in acquiring companions through dynamical exchanges \citep{ziosi2014}, because they are more massive than other stars in star clusters: exchanges favour the formation of more massive binaries (which are more energetically stable, \citealt{hills1980}). Because of their formation mechanism, exchanged BBHs are significantly more massive than both isolated BBHs and original BBHs (cf. \citealt{dicarlo2019}). 

Moreover, some of the exchanged BBHs contain BHs born from mergers between two or more stars. These BHs can be significantly more massive than BHs born fom single stars: the maximum BH mass in the simulations by \cite{dicarlo2019} is $\sim{}440$ M$_\odot$. Such massive BHs born from the merger of two or more stars are initially single objects, but they can acquire companions through dynamical exchanges.

\subsection{Treatment of spins and redshift}

\subsubsection{Spins}
\label{Sec_description_spin}
The initial magnitude and direction of BH spins is still a matter of debate (see e.g. \citealt{millermiller2015} for a review). Overall, the dependence of the spin magnitude of a BH on the spin magnitude of the progenitor star (or stellar core) is largely unknown. 

As for the direction of the spins, most binary evolution processes in isolated binaries tend to favour the alignment of stellar spins with the orbital angular momentum of the binary. %
 In isolated binaries, supernova kicks are the leading mechanism to substantially tilt the spin axes with respect to the orbital plane \citep{kalogera2000}.
In star clusters, dynamical exchanges tend to reset the memory of the initial binary spin. Thus, we expect the spins of exchanged BBHs to be isotropically distributed.
The spin direction of original BBHs in star clusters
is expected to fall somewhere in between because, on the one hand, these BBHs participate in the dynamical evolution of the star cluster (thus, dynamical encounters can affect the initial spin orientation), while on the other hand they form from the evolution of stellar binaries (thus, binary evolution processes tend to realign the spins).

\setlength{\tabcolsep}{3pt}
\begin{table}[h!]
\begin{center}
\begin{tabular}{cccc}
\hline
Model & Rms  & Orientation  & BBH sample \\
\hline
LSA  & 0.1   & aligned      & Isolated BBHs, Original BBHs\\
LSI  & 0.1   & isotropic    & Exchanged BBHs
\vspace{0.3cm} \\
HSA  & 0.3   & aligned      & Isolated BBHs, Original BBHs \\
HSI  & 0.3   & isotropic    & Exchanged BBHs \\
\hline
\end{tabular}
\end{center}
\begin{flushleft}
\caption{Summary of our spin models. We implement two prescriptions (L: low; H: high) of the spin magnitude by varying the root mean square of their Maxwellian distribution, and two  prescriptions for the spin orientations (A: aligned; I: isotropic). Exchanged BBHs are always assumed to have isotropic spins (LSI or HSI), while isolated BBHs and original BBHs are assumed to have aligned spins (LSA or HSA).}
\label{tab:table1}
\end{flushleft}
\end{table}
Given these considerable uncertainties on both spin magnitude and direction, we decided not to embed detailed spin models in our population synthesis and dynamical simulations. Spins are added to our simulations in post-processing, assuming simple toy models.   
Dimensionless spin magnitudes $a$ (defined as $a=|J|\,{}c/G\,{}m_{\rm BH}^2$, where $J$ is the BH spin, $c$ is the speed of light, $G$ is the gravitational constant and $m_{\rm BH}$ is the mass of the BH)
 are randomly drawn from a Maxwellian distribution with root mean square equal to $0.1$. With this choice, the median spin is $a\sim{}0.15$ and the distribution quickly fades off  for $a>0.4$. Hereafter, we refer to this model as ``low-spin'' (LS). 
We assume this distribution for spin magnitudes because the results of the first two aLIGO/aVirgo runs disfavour distributions with large spin components aligned
(or nearly aligned) with the orbital angular momentum \citep{abbottO2popandrate}. %

For comparison, we also consider a second rather extreme case, in which spin magnitudes are drawn from  a Maxwellian distribution with root mean square equal to $0.3$ (we reject spin magnitudes $a>0.998$). With this choice, the median spin is $a\sim{}0.46$. Hereafter, we refer to this model as ``high-spin'' (HS).

Regarding spin orientations, we assume that BH spins in both isolated BBHs and original BBHs are perfectly co-aligned with the orbital angular momentum of the binary (see e.g.~\citealt{rodriguez2016c}): there are large uncertainties on the kicks imparted on newly formed BHs, but recent work \citep{gerosa2018} shows that the fraction of BBHs with negative effective spins is at most $\sim 20\%$.
Moreover, we neglect the effect of dynamical perturbations on original BBHs because their main properties are similar to isolated BBHs: they have nearly the same chirp mass, mass ratio and eccentricity distribution \citep{dicarlo2019}.

Finally, BH spins in exchanged BBHs are randomly drawn isotropically over a sphere. Our assumptions for the spin models are summarized in Table~\ref{tab:table1}.

\subsubsection{Redshifts}
\label{subsecredsfhit}

The redshift parameter was not computed self-consistently in the set of astrophysical simulations generated for this study, because we only consider a fixed value for the metallicity, and the stellar metallicity is a crucial ingredient of redshift evolution \citep{mapelli2017}. A self-consistent redshift evolution will be included in future work %
(Baibhav et al., in preparation). 

Here we opted for excluding redshift information from our statistical analysis (cf. Sec.~\ref{sec_4}). However, we still need to prescribe a redshift probability distribution function to estimate selection effects. 
As a simple toy model, we assume that the redshift is distributed  uniformly in comoving volume and source-frame time, i.e.
\begin{equation}
p(z) \propto \dfrac{1}{1 + z} \dfrac{\text{d} V_{c}}{\text{d} z}. 
\label{sec_dist_z}
\end{equation}
We consider redshifts in the range $z \in [0,2]$ for both second- and third-generation detectors, postponing more accurate modelling to future work.

\subsection{Catalog distributions} 
\label{sec_pres_catalog}

\begin{figure}
\plotone{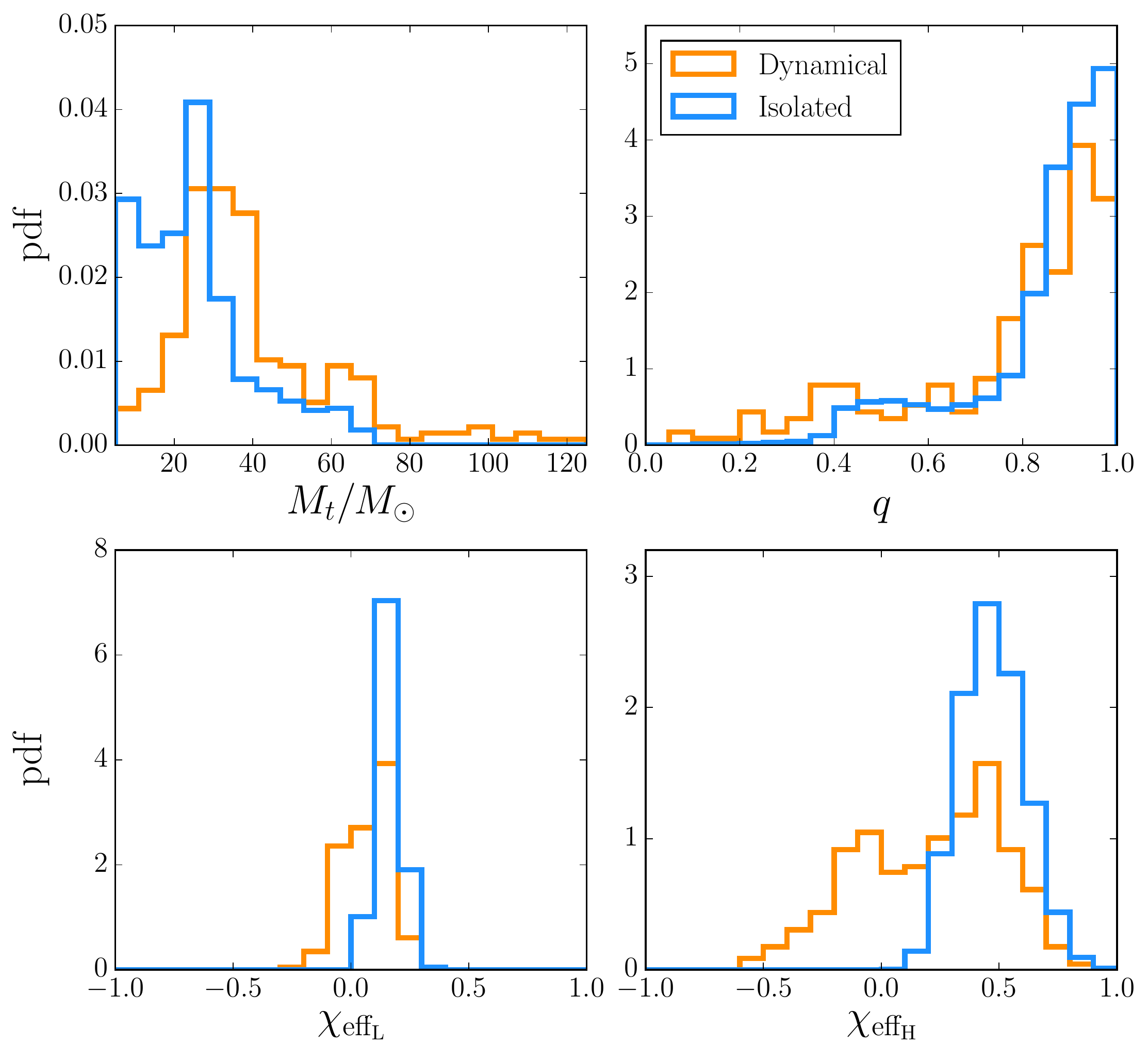}
\caption{Astrophysical population of merging BBHs from dynamical (orange) and isolated (blue) formation channels as presented in Section \ref{sec_2}. We show distributions of total mass $M_{t}$, mass ratio $q$ and effective spin parameters for low-spin ($\chi_{\text{eff}_{\text{L}}}$) and high-spin ($\chi_{\text{eff}_{\text{H}}}$) cases.}
\label{astro_pop}
\end{figure}

In Figure \ref{astro_pop}, we present our distributions from both dynamical (orange) and isolated (blue) catalogs. We plot  distributions corresponding to total mass ($M_{t}$),  mass ratio ($q$) and  effective spins for both the LS ($\chi_{\text{eff}_{\text{L}}}$) and HS ($\chi_{\text{eff}_{\text{H}}}$) models. %

The isolated model allows for BBHs with total mass in the range $M_{t} \in [5,70] M_{\odot}$. On the other hand, the dynamical case presents massive BBHs with $M_{t} > 70 M_{\odot}$ formed via dynamical interactions (exchanged BBHs). The dynamical model predicts BBHs with $q<0.2$, which are not present in the isolated BBH catalogs. The physical reasons for the difference between the maximum mass of isolated BBHs and dynamical BBHs are throughly explained in \cite{dicarlo2019}. Here we summarize the main ingredients. First, BBHs with $M_{t} > 70 M_{\odot}$ form even in our isolated binaries, but they are too wide to merge within a Hubble time (see \citealt{giacobbo2018a}). In the dynamical simulations, these massive wide BBHs have a chance to shrink by dynamical interactions and to become sufficiently tight to merge within a  Hubble time. Secondly, massive single BHs (with mass $\gg{}30$ M$_\odot$) form from collisions between stars (especially if one of the two colliding stars has already developed a Helium core). If these massive single BHs are in the field, they likely remain alone, while if they form in the core of a star cluster, they are very efficient in acquiring new companions through exchanges.

By construction,
the isolated scenario only contains binaries with $\chi_{\text{eff}}>0$, %
while dynamically-formed BHs are found with both positive and negative values for $\chi_{\text{eff}}$ (with a preference for positive values).

\section{Statistical analysis}
\label{sec_3}

\subsection{GW data analysis}

\subsubsection{Detection probability}  
\label{sec_det_prob}
We estimate selection effects using the semi-analytic approach of \citealt{finn1992} (cf. also \citealt{dominik2015,2017arXiv170908079C,taylor2018}).
We associate a detection probability $p_{\text{det}}(\lambda) \in [0,1]$ to any given GW source with parameters $\lambda$. A source is detectable if its signal-to-noise ratio (SNR)
\begin{equation}
  \rho = 4 \int_{0}^{+\infty} \dfrac{| \tilde{h}(f) |^{2}}{S_{n}(f)} \text{d}f
  \label{SNR_def}
\end{equation}
exceeds a given threshold $\rho_{th}$, with $\tilde{h}(f)$ being the gravitational waveform in the frequency domain and $S_{n}(f)$ the one-sided noise power spectral density of the detector. To compute $\tilde{h}(f)$, we have used the IMRPhenomD model \citep{khan2015}, that is a phenomenological waveform model describing the inspiral, merger and ringdown of a non-precessing BBH merger signal.
We consider the noise power spectral density curves corresponding to both the aLIGO (\cite{abbott2018}) and the Einstein Telescope (ET,\cite{evans2016}) at their design sensitivity. We implement a single-detector SNR threshold $\rho_{th}=8$, which was shown to be a good approximation of more complex multi-detector analysis based on large injection campaigns (see \citealt{abadie2010,abbott2016_2,wysocki2018} for more details). Both waveforms and detector sensitivities were generated using \textsc{pycbc} \citep{canton2014,usman2015}.

For each binary in our catalogs, we estimate the optimal SNR, $\rho_{\text{opt}}$, using Eq.~\eqref{SNR_def}. This corresponds to a face-on source located overhead with respect to the detector. The SNR of a generic source is given by $\rho=\omega\times \rho_{\text{opt}}$, where $\omega$ encapsulates all the dependencies on sky-location, inclination and polarization angle \citep{finn1992,finn1996}. A source located in a blind spot of the detector yields a value of $\omega=0$, while an optimally oriented source has $\omega=1$. The probability of detecting a source is then expressed as
\begin{eqnarray}
p_{\text{det}}(\lambda) &=& \mathbb{P}(\rho \geq \rho_{\text{thr}}) \\
&=&\mathbb{P}(\omega \geq \rho_{\text{thr}} / \rho_{\text{opt}})  \\
&=& 1 - F_{\omega}(\rho_{\text{thr}} / \rho_{\text{opt}}),
\end{eqnarray}
where $F_{\omega}$ is the cumulative distribution function of $\omega$. This function was computed via Monte Carlo methods as implemented in the python package \textsc{gwdet} \citep{gwdet_ref}. The function $F_{\omega}$ is set explicitly to 1 for $\rho_{\text{opt}} < \rho_{\text{thr}}$, which gives the expected detection probability $p_{\text{det}}(\lambda) = 0$ for events which are too quiet to be observed.

\subsubsection{Measurement errors}
\label{sec_measurement_errors}

The noise contained in the data $d$ of a GW detector results in errors on the measurement of the parameters $\lambda$ of a GW source. From a Bayesian point of view, these errors are fully described by the posterior distribution $p(\lambda | d)$. We make use of posterior distributions for the first 10 GW events publicly released by  \cite{abbottO2}. We also generate mock observations from our catalogs to forecast future scenarios with a growing number of events. In this case, running a full injection campaign to estimate measurement errors would be computationally too expensive and out of the scope of this study. For simplicity, we approximate posterior distributions with simple Gaussians \citep{gerosa2017,farr2017}
\begin{equation}
p(\lambda_{i}) = \mathcal{N}(\overline{\lambda_{i}},\sigma_{i}).
\label{def_error}
\end{equation} 
The mean $\overline{\lambda_{i}}$ are obtained by first extracting a value $\lambda_{i}^{T}$ from our astrophysical models. We then use prescriptions \footnote{Compared to \cite{mandel2017}, we replace $12/\rho\to 8/\rho$ to account for the fact that in this paper we use sigle-detector SNRs as proxies for detection; cf. e.g. \cite{abbottO2popandrate}.} by \cite{mandel2017} (see also  \citealt{stevenson2017b,powell2019}): injected chirp mass $M_c$ and symmetric mass ratio $\eta$ are given by
\begin{eqnarray}
\overline{\mathcal{M}_{c}} &=& \mathcal{M}^{T}_{c} \left[ 1 + \alpha \dfrac{8}{\rho}r_{0} \right],  \\
\overline{\eta} &=& \eta^{T} \left[ 1 + 0.03 \dfrac{8}{\rho}r_{0} \right],  \\
\overline{\chi_{\text{eff}}} &=& \chi_{\text{eff}}^{T} + \beta \dfrac{8}{\rho} r_{0},
\end{eqnarray}
where $r_{0} \sim \mathcal{N}(0,1)$, $\beta = 0.1$ and $\alpha$ takes values of $0.01$, $0.03$ and $0.1$ for $\eta^{T}>0.1$, $0.1 > \eta^{T} > 0.05$ and $\eta^{T}< 0.05$ respectively, and we convert $(\overline{\mathcal{M}_{c}},\overline{\eta}) \to (\overline{M_{t}},\overline{q})$. Finally, the values for the standard deviations $\sigma^{i}$ are set using the same prescriptions as \cite{mandel2017} given in the previous set of equations.

Measurement errors for ET are obtained by rescaling the aLIGO results using the SNR
\begin{equation}
\sigma_{\text{ET}} = \sigma_{\text{aLIGO}} \dfrac{\rho_{\text{aLIGO}}}{\rho_{\text{ET}}}\,,
\end{equation}
as expected in the large-SNR limit \citep{poisson1995}.

\subsection{Bayesian modeling}
\subsubsection{Model rates}

The general expression for the rate of a given model with parameters $\theta$ can be written as
\begin{equation}
\dfrac{\text{d} N }{\text{d} \lambda} (\theta) = N(\theta) p(\lambda | \theta),
\label{expression_rate}
\end{equation}
where $N(\theta)$ is the total number of sources predicted by the model and $p(\lambda | \theta)$ is the normalised model distribution or rate. 

From the catalog of sources presented in Section \ref{sec_pres_catalog}, we approximate the normalised rates $p$ using kernel density estimation (KDE) methods. Gaussian kernels with a bandwidth parameter of $0.05$ on the data set $\lbrace M_{t}, q, \chi_{\text{eff}} \rbrace$ are capable of accurately reproducing the distributions in Figure \ref{astro_pop} for both formation channels and spin models.

\subsubsection{Hierarchical inference}
\label{sec_statistical_description}

Statistical inference is implemented with a standard Bayesian hiearchical model. %
Our analysis is based on the formalism already presented by \cite{loredo2004}, \cite{mandel2018} and \cite{taylor2018}. In a nutshell, the posterior distribution on the model parameters $\theta$ marginalized over $N(\theta)$ is
\begin{equation}
p(\theta | d) \propto p(\theta)  \prod_{k=1}^{N_{\rm det}} \frac{\int p(\lambda|\theta)\,{} p(\lambda|d)/p(\lambda) \,{}{\rm d}\lambda}{\int p_{\text{det}}(\lambda) \,{}p(\lambda|\theta) \,{}{\rm d}\lambda}\,,
\label{hiermdoel}
\end{equation}
where $N_{\rm \text{det}}$ is the number of entries in the detection catalog, $p(\lambda|\theta)$ describes the astrophysical model, $p(\theta)$ is the prior on each astrophysical model, $p(\lambda| d)$ is the posterior of an individual GW event, $p(\lambda)$ is the prior used in the single-event analysis, and $p_{\rm det}(\lambda)$ describes selection effects.

If the posterior $p(\lambda|d)$ is provided in terms of Monte-Carlo samples ${\lambda_i}$, as in \cite{abbottO2}, we can rewrite Eq.~\eqref{hiermdoel} as 
\begin{equation}
p(\theta | d) \propto p(\theta)  \prod_{k=1}^{N_{\rm det}} \frac{\sum_i p(\lambda_i|\theta) /p(\lambda_i)}{\int p_{\text{det}}(\lambda) \,{}p(\lambda|\theta) \,{}{\rm d}\lambda}\,.
\label{posterior_final_LIGO_main}
\end{equation}
For the case of our mock Gaussian posteriors, Eq.~\eqref{hiermdoel} becomes
\begin{equation}
p(\theta | d) \propto p(\theta)  \prod_{k=1}^{N_{\rm det}} \frac{\int p(\lambda|\theta) \mathcal{N}(\lambda_{k},\sigma_{k})/p(\lambda) \,{}{\rm d}\lambda}{\int p_{\text{det}}(\lambda) \,{}p(\lambda|\theta) \,{}{\rm d}\lambda}\,.
\label{post_mock}
\end{equation}

The denominator $\beta(\theta)\equiv \int p_{det}(\lambda) \,{}p(\lambda|\theta) \,{}{\rm d}\lambda$ does not depend on the event parameters $\lambda$, but only on the model $\theta$. In practice, we estimate $\beta(\theta)$ by generating values for $(M_{t},q)$ using rejection sampling from our KDE approximation of the astrophysical rates, and extracting values for the aligned components of the spins and the redshift as described in sections \ref{Sec_description_spin} and \ref{subsecredsfhit}. As the dynamical catalog is formed of both exchanged and originals BBHs, we consider the two sub-populations separately and combine them in the proportion predicted by our dynamical simulations.

\section{Results}
\label{sec_4}

\subsection{Model selection: pure dynamical or pure isolated channel}

We first apply the formalism presented in Section \ref{sec_statistical_description} to the case where the astrophysical model is such that all BBHs are assumed to form only via the isolated or dynamical channels. In this case, we can estimate what model best fits a given set of data by computing the odds ratio,

\begin{equation}
\mathcal{O}_{AB} = \dfrac{p(A|d)}{p(B|d)},
\label{eq_odds_ratio}
\end{equation}
where $A$ and $B$ either stand for isolated or dynamical and $p$ is the model posterior distribution derived at the end of Section \ref{sec_statistical_description}. Values $\mathcal{O}_{AB} \gg 1$ indicate that model A is strongly favoured by the data, while model B is preferred for $\mathcal{O}_{AB} \ll 1$. It is somewhat indicative to relate values of the odds ratio to $\sigma$-levels of Gaussian measurements:
\begin{equation}
\mathcal{O} = \dfrac{1}{1 - \text{erf}(\sigma / \sqrt{2})},
\end{equation} 
where $\text{erf}$ is the error function.

\subsubsection{Mock observations}

We want to assess the number of observations $N_{\text{obs}}$ needed to discrimininate between the two models assuming that one of them is a faithful representation of reality. To understand how each parameter impacts the analysis, we run our statistical pipeline assuming we only measure either $\lbrace M_{t} \rbrace$, $\lbrace M_{t}, q \rbrace $ or $\lbrace M_{t}, q,\chi_{\text{eff}} \rbrace$. In practice, this implies that the integral in Eq.~\eqref{post_mock} is evaluated on the selected variables while marginalising over the others.  Regarding the production of mock data,  we generated $10^{3}$ sets of observations for each value of $N_{\text{obs}}$ in order to produce a statistical estimation on our results. Each of the mock observations was sampled from the model and included in the observation set with probability $p_{\text{det}}$.

\begin{figure*}
\vspace{-3.50mm}
\includegraphics[width=0.49\textwidth, height=5.8cm]{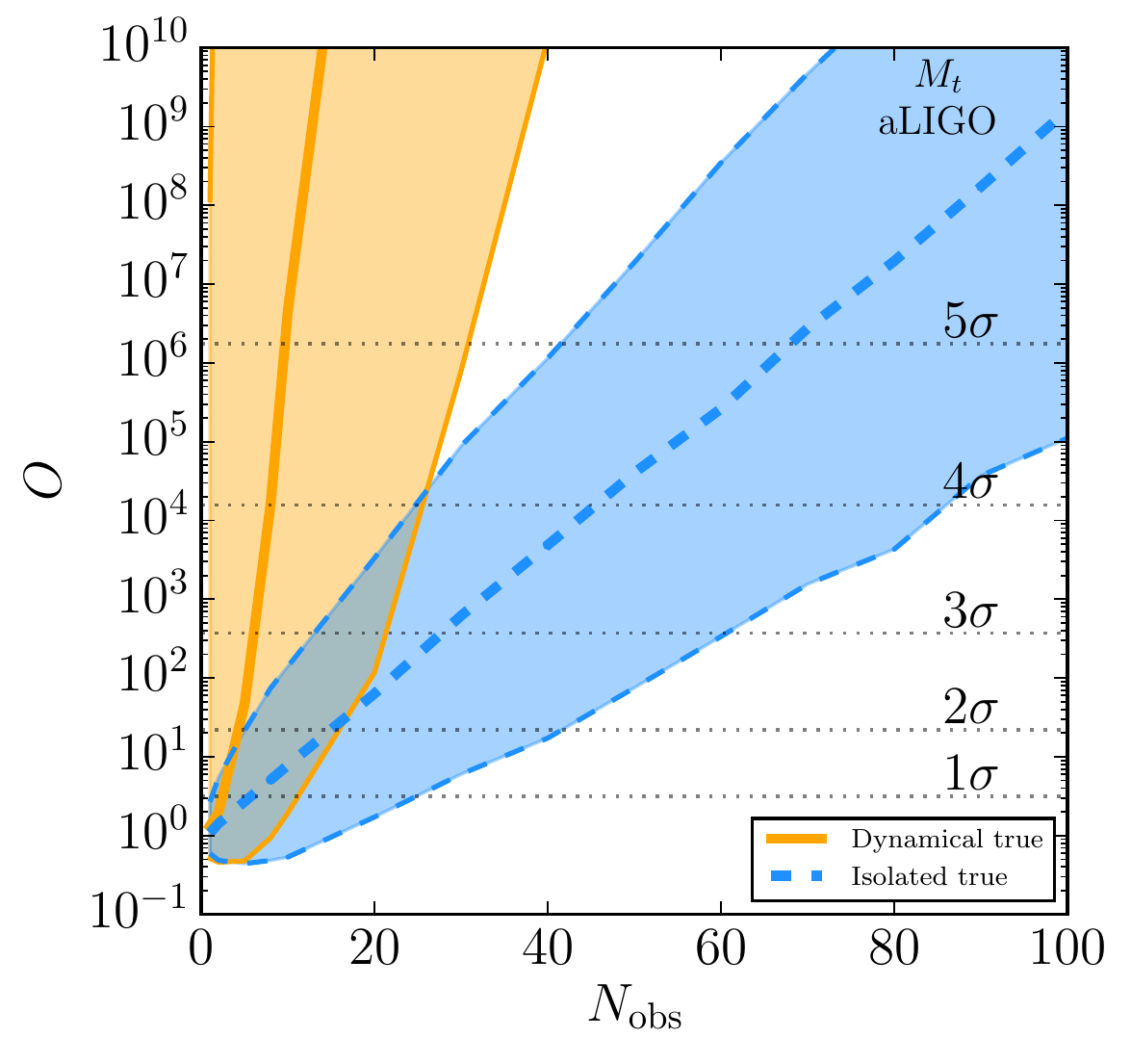}
\includegraphics[width=0.49\textwidth, height=5.8cm]{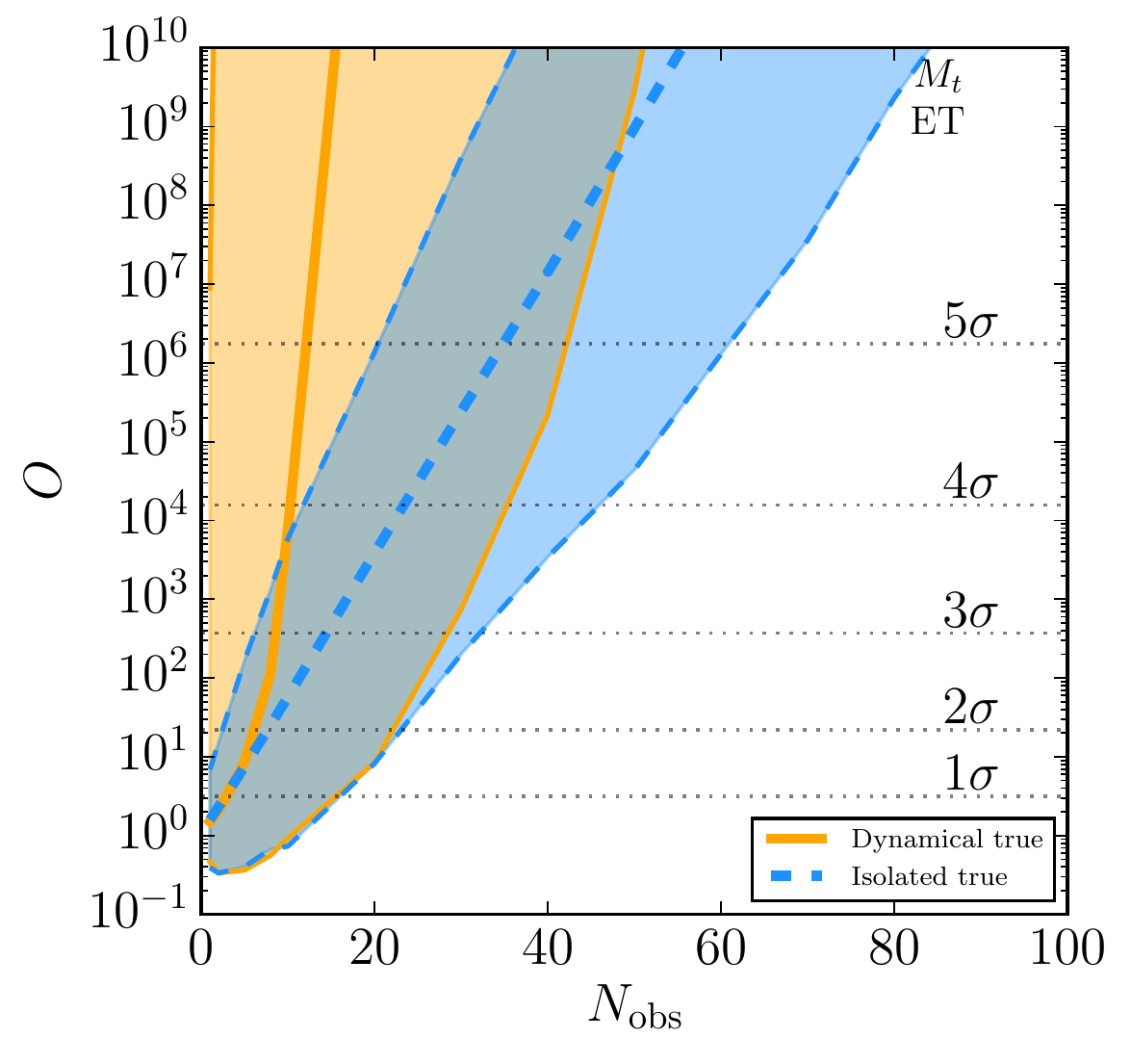}
\vspace{-1.50mm}
\includegraphics[width=0.49\textwidth, height=5.8cm]{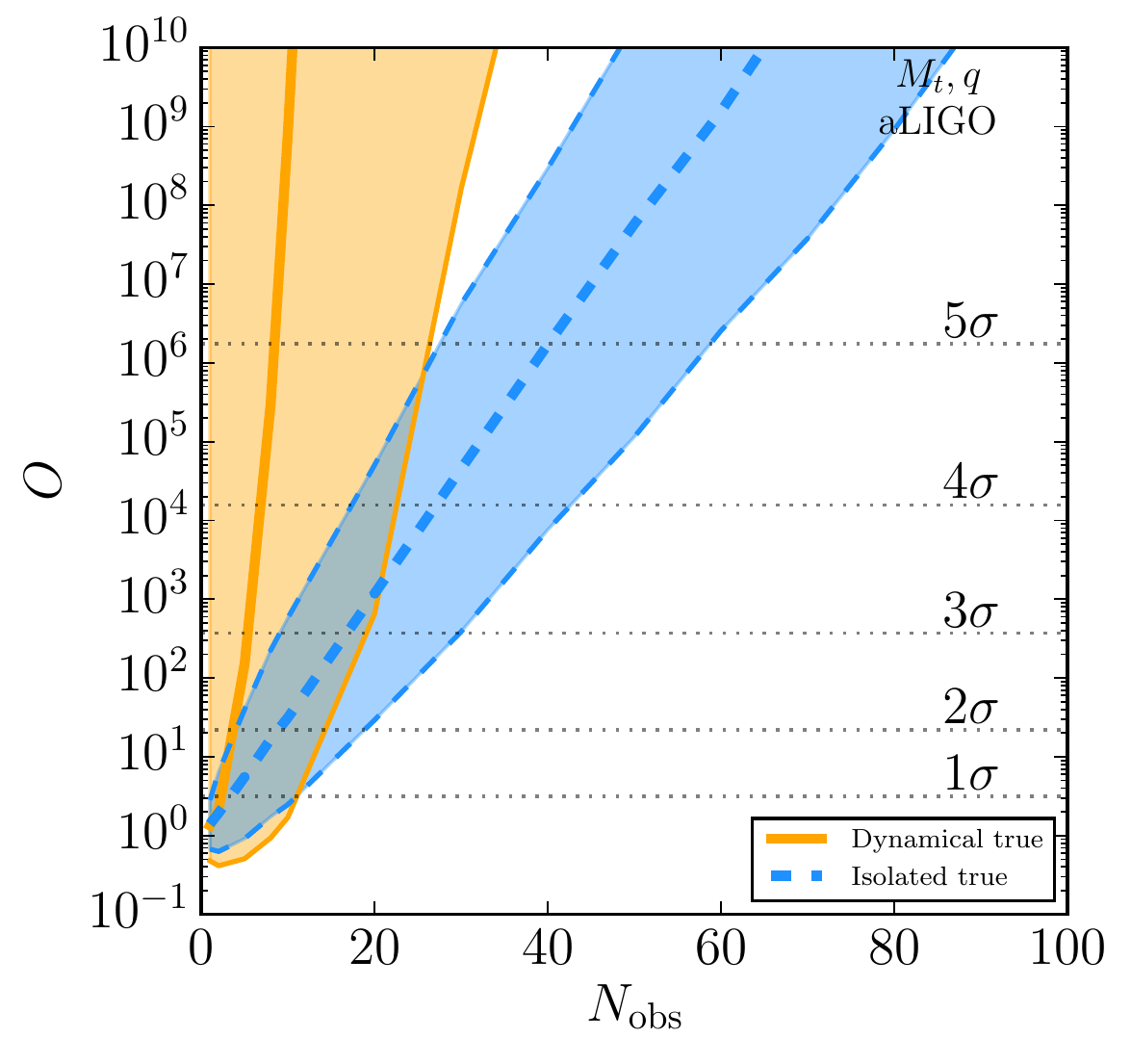}
\includegraphics[width=0.49\textwidth, height=5.8cm]{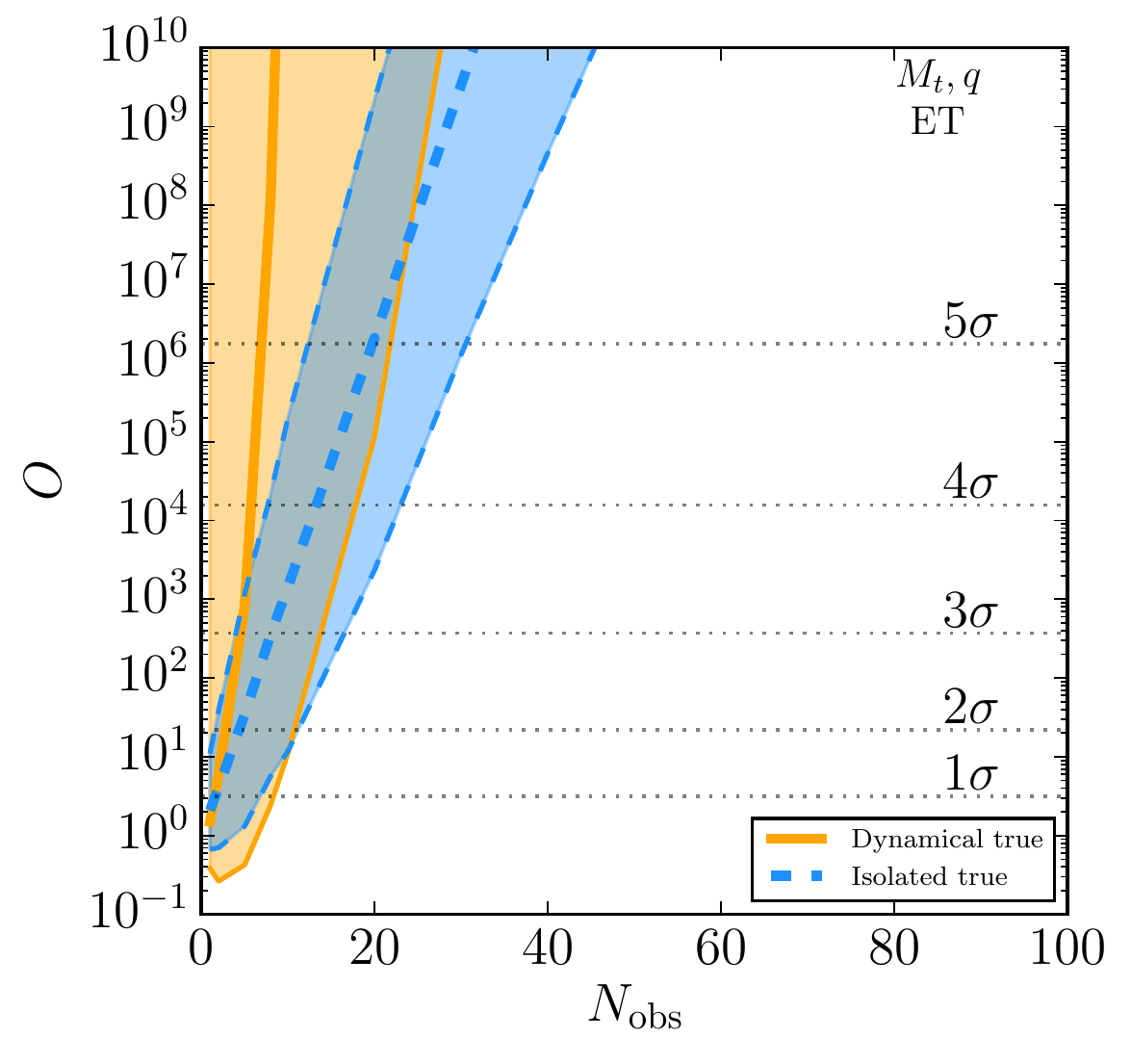}
\vspace{-1.5mm}
\includegraphics[width=0.49\textwidth, height=5.8cm]{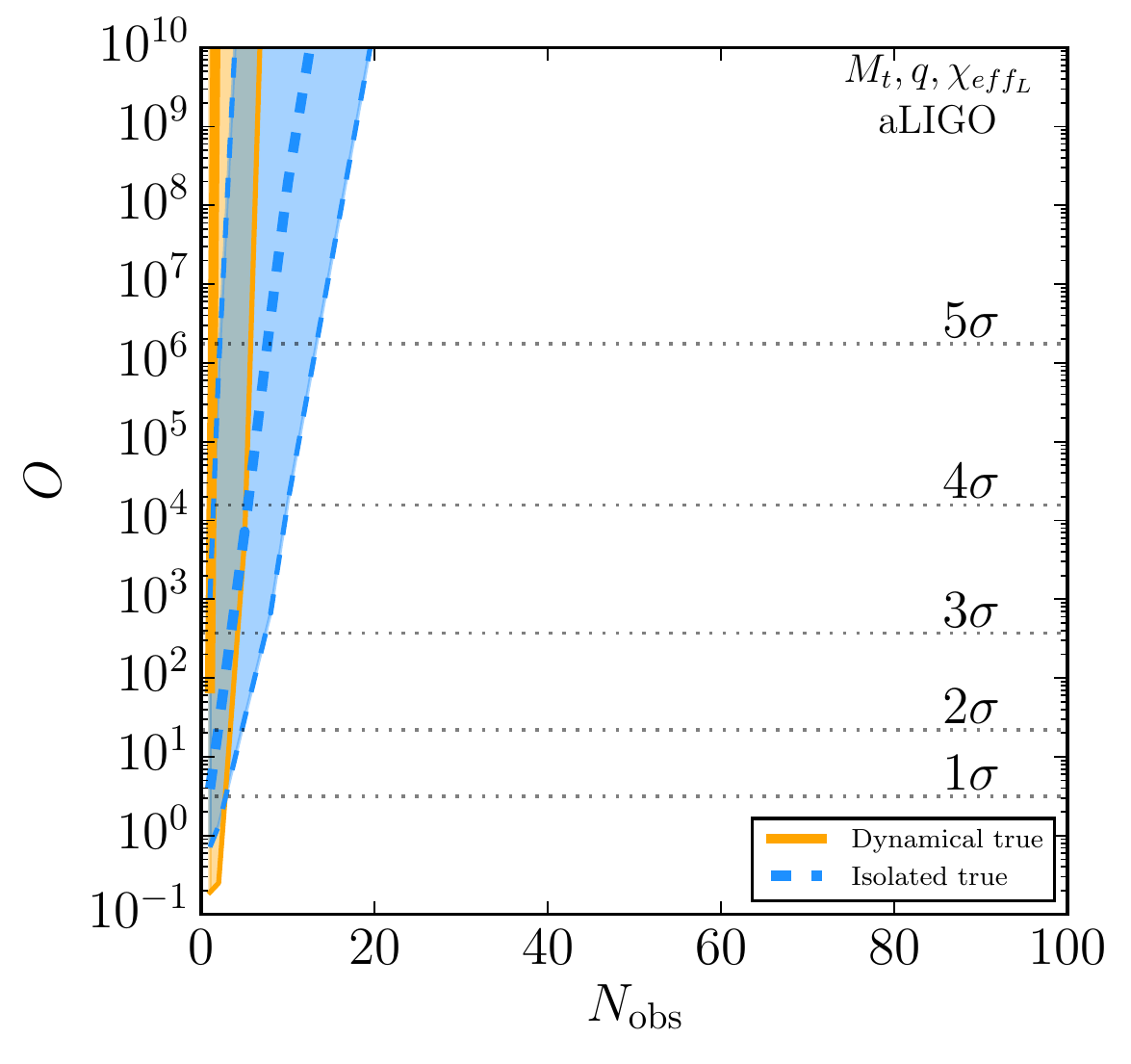}
\includegraphics[width=0.49\textwidth, height=5.8cm]{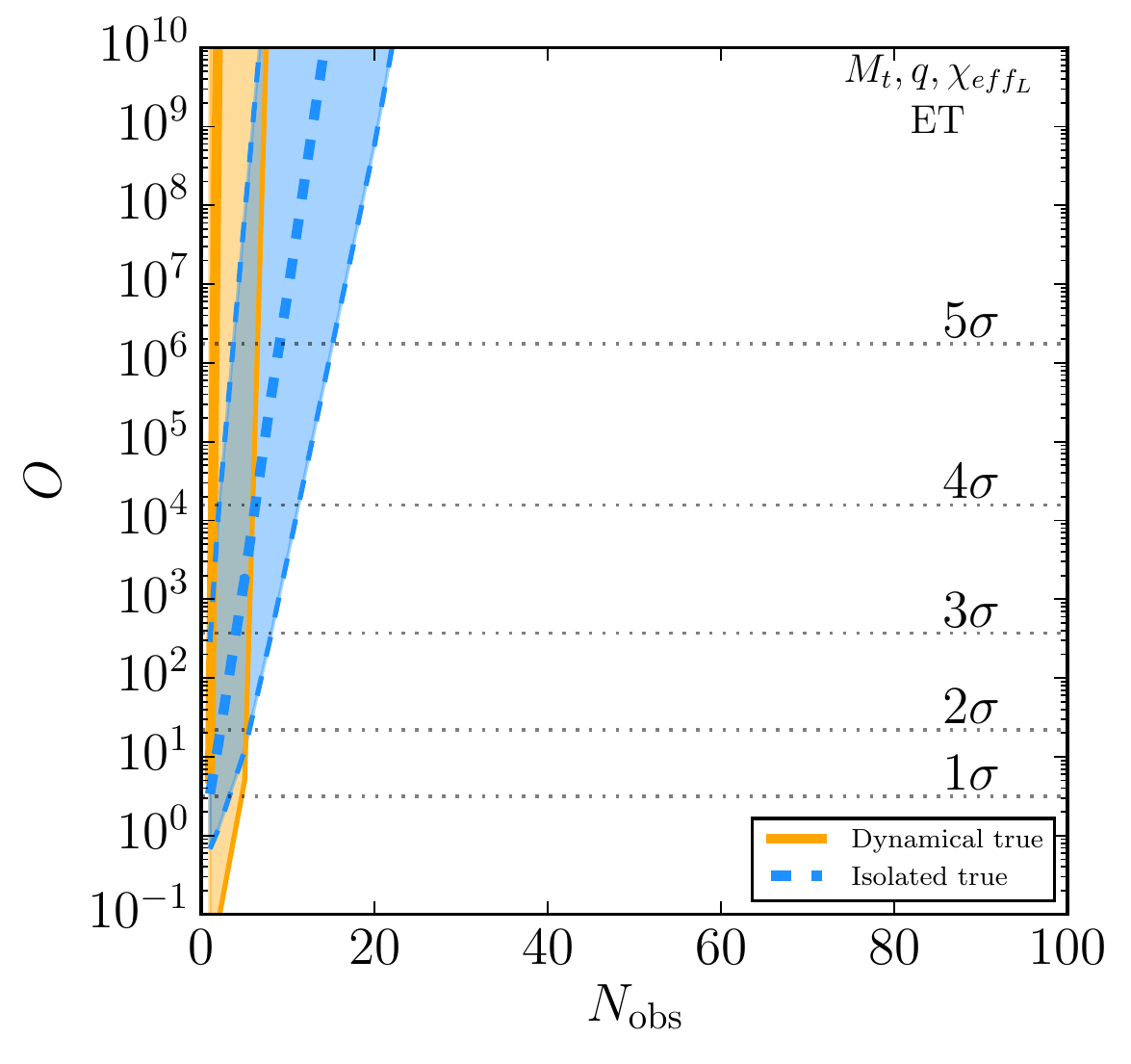}
\vspace{-1.50mm}
\includegraphics[width=0.49\textwidth, height=5.8cm]{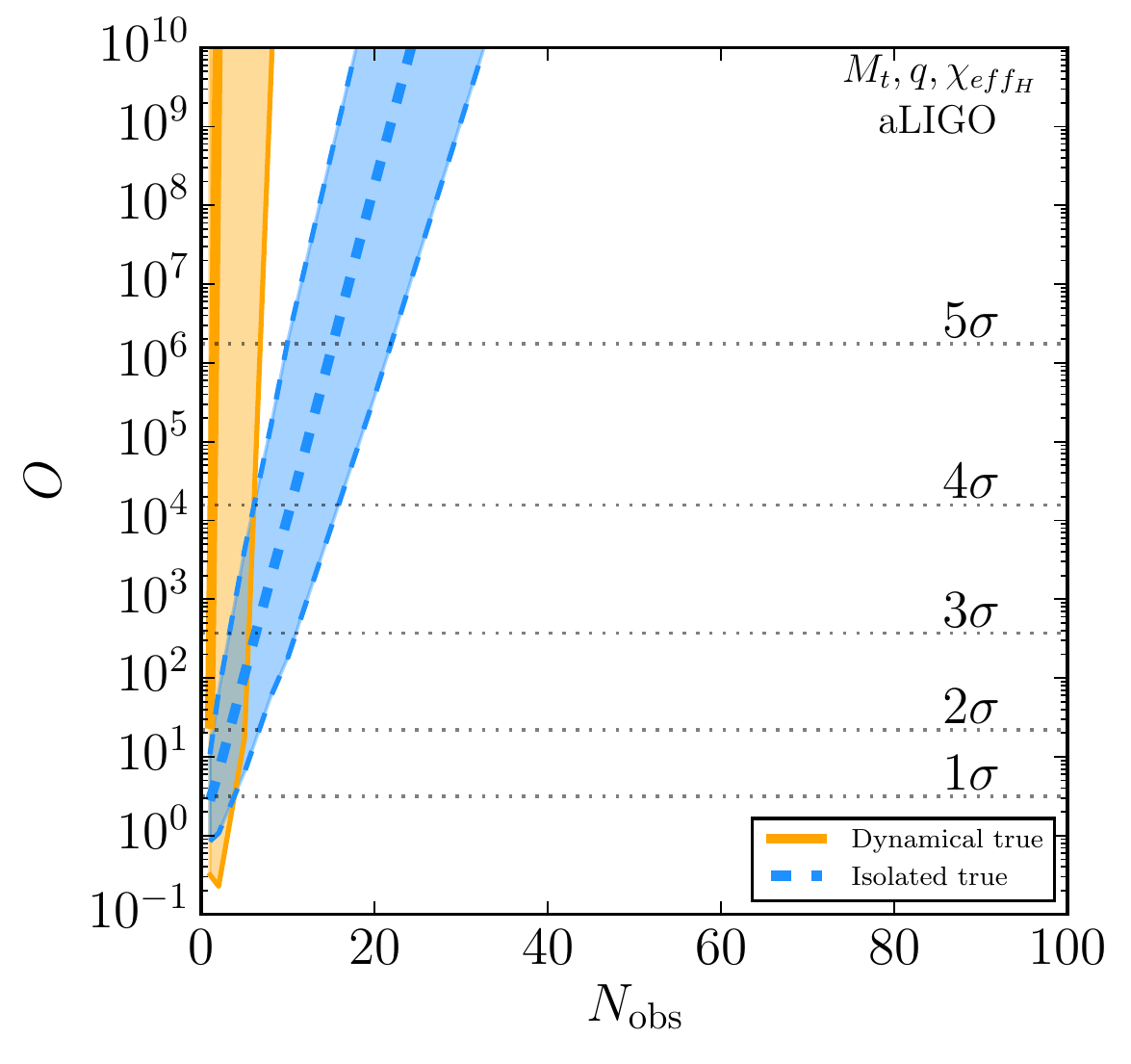}
\includegraphics[width=0.49\textwidth, height=5.8cm]{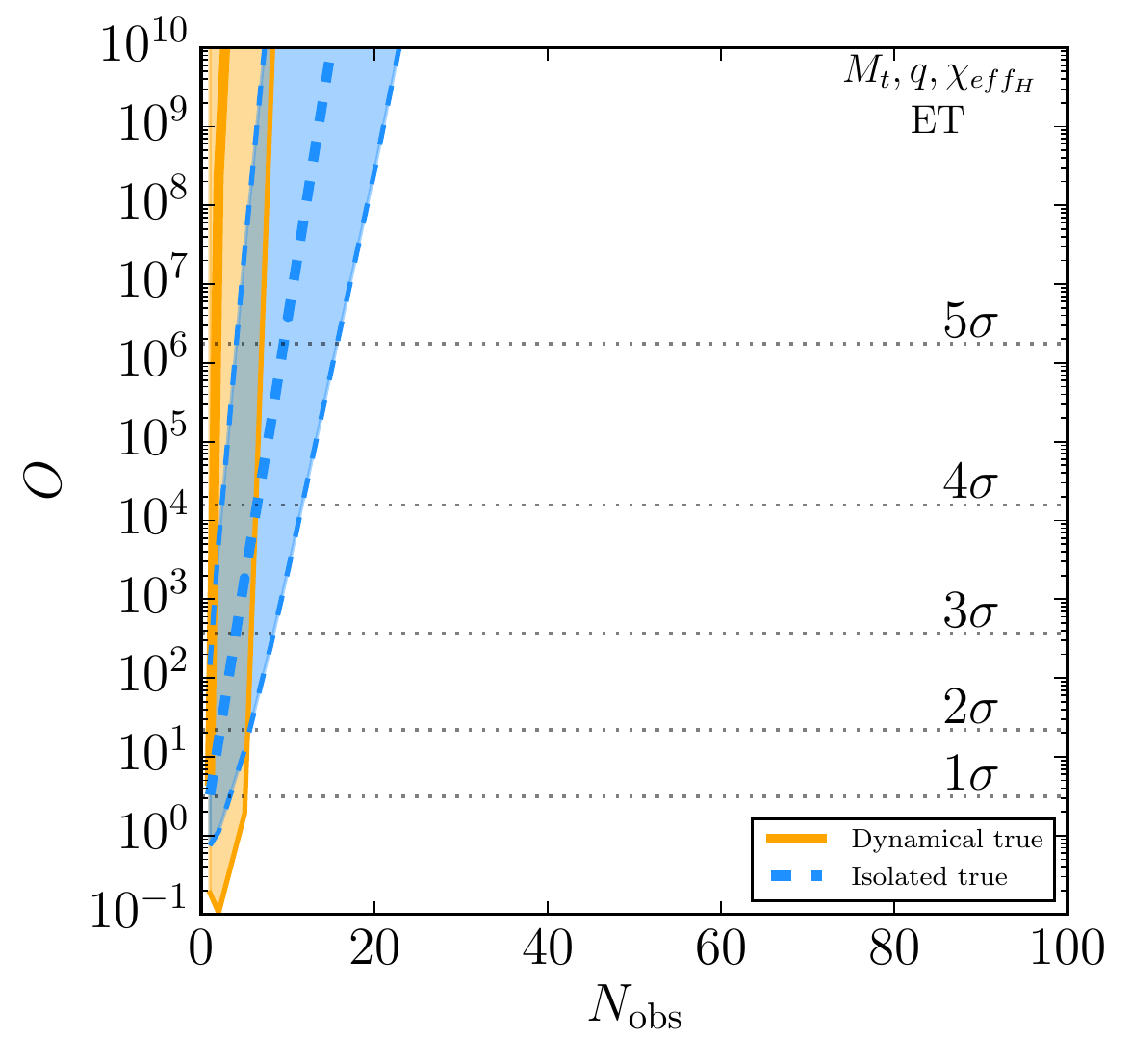}
\vspace{-2.00mm}
\caption{Evolution of the odds ratio as a function of the number of simulated observations generated from the dynamical (orange) and isolated model (blue). The odds ratios are defined such that the posterior of the true model is at the denominator. Equivalent $\sigma$ levels are reported with dotted lines. The four rows correspond to analysis done with $M_{t}$, $q$, $\chi_{\text{eff}_{\text{L}}}$ and $\chi_{\text{eff}_{\text{H}}}$ (from top to bottom); while the left and right panels refer to aLIGO/aVirgo and ET respectively. For each scenario, the thick lines are the median values while the surfaces represent the $90\%$ credible regions.}
\label{fig_pureModel}
\end{figure*}

In Figure \ref{fig_pureModel}, we show values of the odds ratio $O_{AB}$ where the set of observed events is generated from model A, so that $O_{AB}$ is expected to increase as $N_{\text{obs}}$ grows. We present results for observations generated from the dynamical (orange) and isolated (blue) models both for aLIGO/aVirgo (left panels) and ET (right panels). %

Let us focus first on the top row, where only the total mass is considered in the analysis. In the case in which the dynamical model is assumed to be true (orange) with aLIGO/aVirgo, %
 the upper limit of the $90\%$ credible interval presents high values $\mathcal{O} > 10^{10}$ even for small $N_{\text{obs}}$. This is because the isolated model only predicts merging BBHs with $M_{t} < 70 M_{\odot}$. As a result, any observation where the entire support of the total mass posterior distribution is above $70 M_{\odot}$ can only be described by the dynamical model.

Another interesting feature is that the lower bound of the credible interval has values of  $\mathcal{O} > 10^{10}$ starting at $N_{\text{obs}} \sim 40$, which is consistent with our models. In fact, as the number of dynamical BBHs with $M_{t} > 70 M_{\odot}$ in our dynamical catalog is equal to $16$ (out of 229), the probability of not observing any massive BBHs in 50 observations generated from the dynamical model is equal to $p=1.9 \times 10^{-3}$. It is also interesting to see that the results obtained are similar when observing with ET, suggesting that the differences in total mass range of our models is the dominant effect in the analysis.

In contrast, the evolution of the odds ratio for the isolated case presents a steady increase of $\mathcal{O}$ with $N_{\text{obs}}$, as expected.
This indicates that, unlike the dynamical case, there is no strong feature in the total mass spectrum of the isolated model that can drive the odds ratio towards very high values with a few observations. In this case, measurements with ET improve the analysis: the median number of observations yielding $\mathcal{O}=10^{6}$ ($5\sigma$ level) decreases from $65$ with aLIGO down to $\sim 30-35$. 

The second row of Figure \ref{fig_pureModel} presents results obtained when performing the analysis considering both the total mass and the mass ratio. We observe that including the mass ratio helps discriminating faster between the two models for all cases. This is particularly true for ET since in the case where the dynamical model is true, the lower bound of the $90\%$ credible interval reaches the $5\sigma$ level for $N_{\text{obs}} =20$, while we had a value of $N_{\text{obs}} = 40$ for the analysis considering only the total mass. Similarly, in $90\%$ of the cases we find that only $30$ observations generated from the isolated model and observed by ET give values of odds ratio corresponding to a $5\sigma$ level ($60$ observations were needed when considering only $M_{t}$) .

Finally, in the last two rows we present results obtained by simulating measurements with the total mass, the mass ratio and the effective spin parameter, adopting either the low-spin (LS, third row) or high-spin (HS, fourth row)  models. When the dynamical model is true, only a few observations are needed to push the value of the lower bound of the credible interval towards $ \mathcal{O} > 10^{10}$ (similar results were found in \citealt{stevenson2017b}). This results from our assumption for the spin models, as only the dynamical model has support for $\chi_{\text{eff}}<0$, 
meaning that observations of negative values give a full support to the dynamical model. As the errors on the spins are already relatively low, with $\sigma_{\chi_{\text{eff}}} = 0.1$ for aLIGO/aVirgo (see Section \ref{sec_measurement_errors}), we do not see much difference when repeating the analysis with even smaller error for ET. %
Finally, when the isolated model is true, the $5\sigma$ level is attained in $90\%$ of the cases only after $\sim 15-20$ observations for the LS and HS models. Measurements with ET are slightly better and bring the number of necessary observations down to $\sim 10-15$ for both models. Once more, we highlight here that this study was restricted to sources with redshift $z\leq2$, while ET is also expected to see a significant number of sources at high redshift.

\subsubsection{LVC observations}

We now apply the formalism described in the previous Section to aLIGO/aVirgo BBH data from the catalog in \cite{abbottO2}. We have used both the posterior and prior samples for each event released by the LVC, to compute the expression of the posterior in Eq.~\eqref{posterior_final_LIGO_main}. In Table \ref{tab:table2}, we present the values obtained for $\mathcal{O}$ and $\sigma$ for the favoured model under the same parameter configurations presented in the last section. 

For all simulations, the values of the odds ratio are $\mathcal{O}\lesssim10^{4}$, corresponding to $\lesssim4 \sigma$ and favour a dynamical origin. If we only take into account the mass parameters, the analysis done with $\lbrace M_{t},q \rbrace$ gives a value of $\mathcal{O} = 1.7 \times 10^{3}$ (corresponding to $3.4 \sigma$) in favour of the dynamical model. At the fixed metallicity considered here, our isolated model has a hard cutoff on the total mass at $70 M_{\odot}$, which cannot accommodate the presence of GW170729 (but we stress here that this is not the case for all metallicities). 
As a result, if we discard this event from the observed catalog, the odds ratio  drops to $\mathcal{O} = 1.6 \times 10^{1}$ when considering $\lbrace M_{t},q \rbrace$.

Our results depend strongly on the underlying spin model. In the LS case, we get $\mathcal{O}=39$, with slight preference for the dynamical model. This conclusion changes drastically in the HS case, as the value of the odds ratio is now larger than $10^{16}$. This is because in the HS case the dynamical model has a much wider range of possible negative values for $\chi_{\text{eff}}$, extending down to $\chi_{\text{eff}} ~ -0.5$ (compared to $\chi_{\text{eff}} ~ -0.2$ for LS case). As some of the events (namely GW1701014, GW170818, GW170823) have posterior distribution with some support for values of $\chi_{\text{eff}}<-0.25$, the HS dynamical model is significantly favored compared to the HS isolated case. When removing these events from the analysis, we found that the odds ratio is reduced by 10 orders of magnitude down to $2.7 \times 10^{6}$.

\begin{table}[h]
\begin{center}
\begin{tabular}{c|c|c|c}
\hline
Parameters  & Favoured model &  $\mathcal{O}$ & $\sigma$-equivalent  \\
\hline
$M_{t}$  & dynamical & $42$ & 2.3        \\
$M_{t}$,q  & dynamical & $1.7 \times 10^{3}$ & 3.4   \\
$M_{t}$,q,$\chi_{\text{eff}_{\text{L}}}$  & dynamical & $39$ & 2.2   \\
$M_{t}$,q,$\chi_{\text{eff}_{\text{H}}}$  & dynamical & $4.1 \times 10^{16}$ & x   \\
\hline
\hline
$M_{t}$,q [1]  & dynamical & $16$ & 1.9   \\
$M_{t}$,q,$\chi_{\text{eff}_{\text{H}}}$ [2]  & dynamical & $2.7 \times 10^{6}$ & 5.1   \\
\hline
\end{tabular}
\end{center}
\caption{\label{tab:table2} Summary of the results obtained when running Bayesian model selection on the set of 10 BBHs observed by aLIGO/aVirgo \citep{abbottO2}. We run our statistical analysis varying the set of parameters considered. The configurations $[1]$ and $[2]$ correspond to cases where we respectively exclude GW170729 and the set of events (GW1701014, GW170818, GW170823) from the analysis. Among these models, we prefer a dynamical origin mainly because of the hard cuts in the mass and spin distributions (cf. section~\ref{discussionsec}). %
} \leavevmode
\end{table}

\subsection{Mixture model}

In the previous section, we have shown results obtained when comparing the two models assuming that one of them is true. While this pointed out interesting features of our models, it is probably unrealistic to assume that {\emph{all}} merging BBHs only come either from the dynamical or the isolated channel. In this section, we assume that the population of merging BBHs comes from a mixture of the two models parametrised by a ``mixing fraction'' $\mathlcal{f}$ %
such that
\begin{equation}
r_{\text{MM}}(\mathlcal{f}) = \mathlcal{f} \,{}r_{\text{iso}} + (1 - \mathlcal{f})\,{} r_{\text{dyn}},
\end{equation} 
where $r_{\text{MM}}(\mathlcal{f})$ are the rates of the mixture model and $r_{\text{iso}},\,{}r_{\text{dyn}}$ are the rates of the isolated and dynamical models, respectively. Similarly, the detection efficiency of the mixed model is also given by
\begin{equation}
\beta_{\text{MM}}(\mathlcal{f}) = \mathlcal{f} \,{}\beta_{\text{iso}} + (1 - \mathlcal{f}) \,{}\beta_{\text{dyn}}
\end{equation}
where  $\beta_{\text{iso}},\,{}\beta_{\text{dyn}}$ are the detection efficiencies (cf. Sec.~\ref{sec_statistical_description}) for the isolated and dynamical models, respectively.

We want to estimate the posterior distribution of the mixing fraction $\mathlcal{f}$. Once again, we have considered two different cases with either fiducial or aLIGO/aVirgo data. In both cases, we have used a Metropolis-Hastings Monte Carlo algorithm to estimate the posterior distribution. We find that a Gaussian jump proposal with standard deviation $\sigma=0.5$ is sufficient to have good results and convergence when running $10^{6}$ iterations chains.

\subsubsection{Mock observations}

We generate mock observations assuming a ``true'' mixing fraction $\mathlcal{f}_{T} = 0.7$. To generate $N_{\text{obs}}$ events from a mixed model, we first generated $N_{\text{obs}}$ events both for isolated and dynamical models. For each entry of the mixed model, we draw a random number $\epsilon \in \mathcal{U}[0,1]$ and associate an event from the pre-processed isolated (dynamical) set if $\epsilon < \mathlcal{f}_{T}$ ($\epsilon > \mathlcal{f}_{T}$). 

\begin{figure*}
\plotone{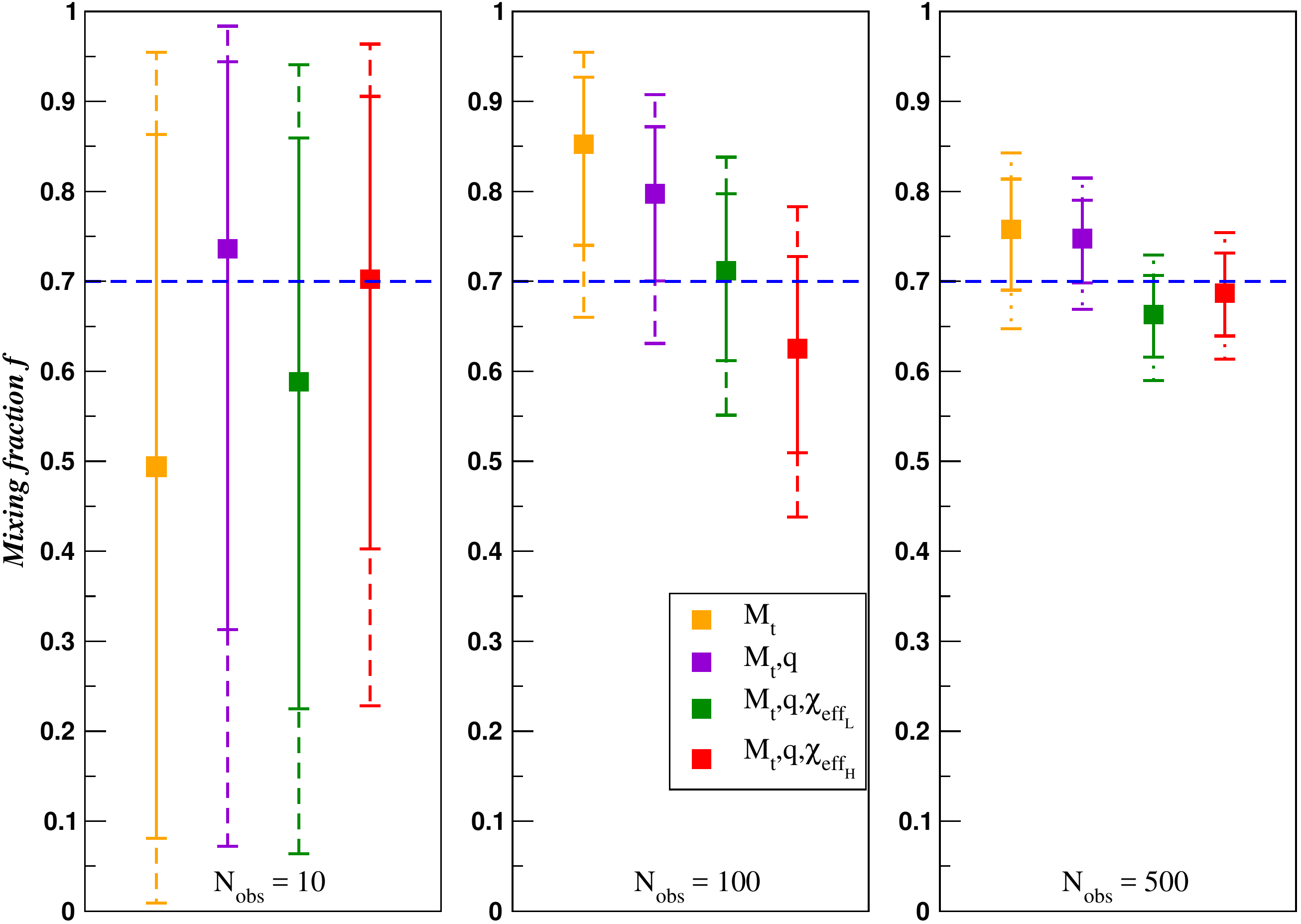}
\caption{Median (squares), $90\%$ (thick lines) and $99\%$ (dashed lines) credible intervals for the mixing fraction posterior distribution as a function of the set of parameters used for the analysis. $\mathlcal{f}=0$ ($\mathlcal{f}=1$) corresponds to the pure dynamical (isolated) model. In all cases, the fiducial data set were generated from a mixed model with mixing fraction $\mathlcal{f}_{T} = 0.7$ and with number of observations $N_{\text{obs}} = 10, 100, 500$ (from left to right). }
\label{fig:mcmc_gen_obs}
\end{figure*}

In Figure \ref{fig:mcmc_gen_obs}, we report the values of the medians (square), $90\%$ (straight line) and $99\%$ (dashed line) credible intervals for a set of mock observations with $N_{\text{obs}}=\lbrace 10, 100, 500 \rbrace$.  The values correspond to the current number of events ($N_{\text{obs}} = 10$), an optimistic prediction for O3 ($N_{\text{obs}} = 100$) and a high-statistic case ($N_{\text{obs}} = 500$). For simplicity, we restrict this study to the aLIGO/aVirgo detector case.  For each set of observations we perform the analysis  with the combination of parameters $\lbrace M_{t} \rbrace$ (orange), $\lbrace M_{t},q \rbrace$ (purple), $\lbrace M_{t},q,\chi_{\text{eff}_{\text{L}}} \rbrace$ (green) and $\lbrace M_{t},q,\chi_{\text{eff}_{\text{H}}} \rbrace$ (red). 

For any set of parameters, we observe a reduction of the width of the credible intervals for higher values of $N_{\text{obs}}$, as expected. In fact, for $N_{\text{obs}} = 10$ the $99\%$ credible interval spans almost the entire range of values for $\mathlcal{f}$, while for $N_{\text{obs}} = 500$ it is reduced down to $0.2$ for $M_{t}$ and only $\sim 0.15$ when including $q$ and $\chi_{\text{eff}}$ in the analysis. Another feature shown in Figure \ref{fig:mcmc_gen_obs} is that the width of the credible interval gets smaller when including more parameters, which is expected as more parameters provide more constraints on the model selection analysis.

In conclusion, Figure \ref{fig:mcmc_gen_obs} suggests that already with $100$ detections (optimistic scenario for the end of O3), we can constrain the value of the fractional errors on the mixing fraction to an interval smaller than $0.5$ using $\lbrace M_{t},q\rbrace$.  This result is in agreement with previous studies \citep{vitale2017,stevenson2017b,zevin2017,talbot2017}. Furthermore, with even higher statistics of a few hundred detections, the fractional errors on the mixing fraction will go down to $20\%$, if we consider only the mass parameters. The inclusion of the effective spin parameter reduces this value even further down to $10\%$. Finally, we have repeated this study for a value of $\mathlcal{f}_{T} = 0.3$, and found similar predictions.

\subsubsection{LVC observations}

\begin{figure}
\plotone{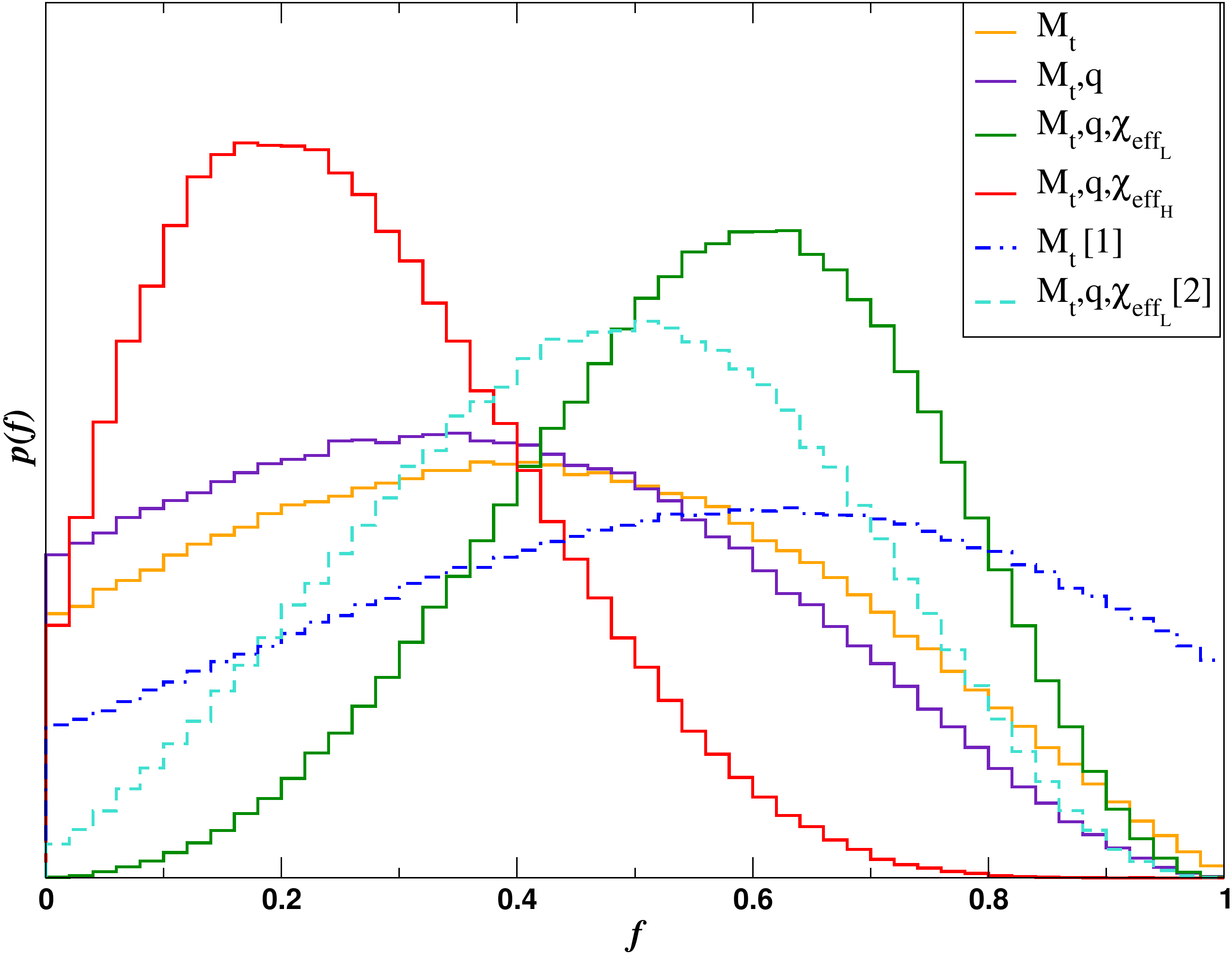}
\caption{Posterior distribution inferred from a $10^{6}$ MCMC chain ran on the LVC observations. The legends $[1]$ and $[2]$ correspond to cases where the events GW170729 and GW151226 were respectively excluded from the analysis. $\mathlcal{f}=0$ ($\mathlcal{f}=1$) corresponds to the pure dynamical (isolated) model. The set of parameters considered for the analysis are indicated in the legend.}
\label{ligo_data_mixture}
\end{figure}

In this section, we describe the results obtained when applying our mixed model analysis to aLIGO/aVirgo data \citep{abbottO2}. Figure \ref{ligo_data_mixture} shows the posterior distributions obtained when doing the analysis using $\lbrace M_{t} \rbrace$ (orange), $\lbrace M_{t},q \rbrace$ (purple), $\lbrace M_{t},q,\chi_{\text{eff}_{\text{L}}} \rbrace$ (green) and $\lbrace M_{t},q,\chi_{\text{eff}_{\text{H}}} \rbrace$ (red).  

First, the two posterior distributions obtained when using the mass parameters are slightly shifted towards values of $\mathlcal{f} < 0.5$ (pure dynamical scenario) with a median value of $0.40$ and $0.36$ for $\lbrace M_{t} \rbrace$ and $\lbrace M_{t},q \rbrace$ respectively. In addition, the upper limit of the $99\%$ credible interval is equal to $0.94$ ($\lbrace M_{t} \rbrace$ case) and $0.90$ ($\lbrace M_{t},q \rbrace$ case), hence excluding our pure isolated scenario. As before, this is due to the fact that some of the detected BBHs have support for $M_{t}>70 M_{\odot}$ while our isolated model strictly prevents the occurrence of these high masses. In particular, for GW170729, the lower $90\%$ credible interval of $M_{t}$ is equal to $74.2 M_{\odot}$. %
For testing purposes, we ran an analysis where the total mass of GW170729 was not included %
(blue curve); in this case, the median was $0.55$ and the upper limit of the $99\%$ credible interval is much higher, with values of $0.99$. As 0.99 is also the $99\%$ upper limit on our prior, this shows that our posterior does not disfavor the isolated scenario any less than the prior. 

Including the effective spin parameter in the analysis has a significant impact on the posterior distributions. In the LS model, the distribution of the mixing fraction is centered towards highers values of $\mathlcal{f}$ (with a median of $0.58$), favouring the isolated scenario. Once again, this result is dominated by a single event, GW151226, that presents an effective spin parameter with a clearly positive median value $\chi_{\text{eff}} = 0.18$ that is not well supported by the dynamical scenario. We re-ran the analysis for the LS model excluding this event (cyan curve) and found that the value of the median is reduced down to a value of $0.48$. It is interesting to highlight that while GW170729 has even higher values of effective spin (median of $\chi_{\text{eff}} = 0.36$), this event gives only limited support to the isolated scenario as the masses are very high (unlike GW151226). In the HS case, the dynamical scenario is favoured with a median value for the mixing fraction equal to $0.27$. This can be understood by the fact that the support for the effective spin parameter in the dynamical case extends between $-0.6$ and $0.7$ (see Figure \ref{astro_pop}), which is the range of all the events currently observed by aLIGO and aVirgo, while the isolated scenario now struggles to capture events with negative values of $\chi_{\text{eff}}$.

In conclusion, our results suggest that O1+O2 LVC data \citep{abbottO2} exclude a pure isolated scenario as described by our population-synthesis simulations, which however include  hard cutoffs on both mass and spin distributions. %
It is important to point out that the width of the posterior distribution is still quite large, and it is in agreement with the results obtained in the last section. Moreover, we stress that the models considered in this paper refer to a single metallicity: at lower progenitor metallicity even the isolated scenario includes higher mass BHs (with $M_t$ up to $80-90$ M$_\odot$, \citealt{giacobbo2018b}). Thus, in a follow-up study (Bouffanais et al., in preparation), we will explore the importance of metallicity and other population-synthesis parameters.

\section{Discussion}
\label{discussionsec}

We have applied a Bayesian model-selection framework to discriminate BBHs formed in isolation versus those those formed in young star clusters. 
Under the assumption that there is only one ``true'' formation channel and for our specific choices of models at metallicity $Z=0.002$, O1+O2 LVC data prefer the dynamical formation channel at $3.4\,\sigma$, if we include in our analysis only the total mass $M_t$ and mass ratio $q$. Similarly, if we adopt a more realistic mixture model, O1+O2 LVC data \citep{abbottO2} exclude a purely isolated scenario as described by our population-synthesis simulations. We stress that this is a very model-dependent statement: our isolated binary formation model contains hard cutoffs in both binary masses ($M_t<70M_\odot$) and spins ($\chi_{\rm eff}>0$), which cannot accommodate some of the events, notably GW170729.%

The effect of the mass cutoff is particularly important. We compared aLIGO data against two models, one with a mass cutoff at $m_1,m_2\sim 35 M_\odot$ (isolated) and one without it (dynamical). Our analysis suggests preference for the model without cutoff. On the other hand, \cite{abbottO2popandrate}  find strong evidence for an upper mass gap starting at $\sim 45 M_\odot$-- this constraint being driven by the lower limit on the largest BH mass in the sample.  Crucially, they do not compare data against astrophysical simulations with a well specified set of assumptions, but rather fit a generic phenomenological population model.
Going forward, this difference highlights the importance of accurately modeling tails of the predicted astrophysical distributions, that despite being responsible for a small number of events, might play a qualitative role in discriminating among different formation channels.

Several caveats need to be discussed about our models. Crucially, the simulations considered here assume a single metallicity $Z=0.1$ Z$_\odot$. Isolated binaries with lower metallicity ($Z\lesssim{}0.01$ Z$_\odot$) simulated with {\sc mobse} end up producing merging BBHs with total mass up to $\sim{}80-90$ M$_\odot$ \citep{mapelli2017,giacobbo2018b}. Thus, the effect of metallicity is somewhat degenerate with the effect of dynamics. We expect a pure isolated scenario to be consistent with O1 and O2 data if we include more metal-poor progenitors (down to $Z\sim{}0.0002=0.01$ Z$_\odot$).

Moreover, the simulations considered in this paper investigate only one model for core-collapse supernovae (from \citealt{fryer2012}) and only one model for pair-instability and pulsational pair-instability supernovae. Furthermore, we assume a specific model for BH natal kicks, which are highly uncertain and can crucially impact the properties of merging systems (e.g. \citealt{belczynskirepetto2016,mapelli2017,barrett2017,wysocki2018,gerosa2018}).  Even the prescriptions for common envelope affect the properties of merging systems significantly (here we consider only one value of $\alpha{}=3$).%

It is worth mentioning here that {\sc mobse} predicts a significant difference between the maximum mass of BHs and the maximum mass of {\emph{merging}} BHs in isolated binaries (see e.g. Figure~11 of \citealt{giacobbo2018b}). From progenitors with metallicity $Z=0.1$ Z$_\odot$ ($Z=0.01$ Z$_\odot$) we form BHs with individual mass up to $\sim{}55$ M$_\odot$ ($\sim{}65$ M$_\odot$), but only BHs with individual mass up to $\sim{}35$ M$_\odot$ ($\sim{}45$ M$_\odot$) merge within a Hubble time through isolated binary evolution. This behaviour is similar to what found from the independent population-synthesis code {\sc sevn} \citep{spera2019} and springs from the interplay between stellar radii and common envelope (see \citealt{giacobbo2018b} and \citealt{spera2019} for more details). Thus, if the progenitor metallicity is $Z=0.1$ Z$_\odot$, merging BBHs from isolated binaries have $M_{\rm t}\leq{}70$ M$_\odot$, while non-merging BBHs from isolated binaries have  $M_{\rm t}\leq{}110$ M$_\odot$.

In contrast, dynamically formed BBHs (especially BBHs formed by dynamical exchange) might be able to merge even if their total mass is larger than $\sim{}70$ M$_\odot$, because common envelope is not the only way to shrink their orbits: dynamical exchanges and three-body encounters also contribute to reducing the binary semi-major axis and/or increasing the orbital eccentricity. This is the main reason why $M_{\rm t}$ is substantially larger for dynamical BBH mergers than for isolated BBH mergers.

In addition, dynamical BBHs can host BHs which form from the collision between two or more stellar progenitors (which is not possible in the case of isolated binaries). In our models, if a BH forms from the evolution of a collision product between an evolved star (with a well-developed Helium or Carbon-Oxygen core) and another star, it might have a significantly larger mass, because a collision product can end its life with a larger total (or core) mass and with a larger envelope-to-core mass ratio than a single star (see \citealt{dicarlo2019} for additional details). This further enhances the mass difference between isolated and dynamical merging BBHs. Moreover, BHs that form from the collisions of two or more stars are also allowed (in some rare cases) to have mass in the pair-instability mass gap, if their final Helium core is below the threshold for (pulsational) pair instability but their Hydrogen mass is larger than expected from single stellar evolution (in our models we impose that the BH mass is equal to the final progenitor mass, including Hydrogen envelope, if there is a direct collapse\footnote{This assumption is still matter of debate given the high uncertainties on direct collapse (see e.g. \citealt{sukhbold2016}).}). From \cite{dicarlo2019} we find that $\lesssim{}2\%$ of all merging BHs born in young star clusters have mass in the pair-instability mass gap. We do not find second-generation BH mergers, because of the low escape velocity from young star clusters \citep{gerosa2019}.

We also stress that the very edges of the pair-instability mass gap are not uniquely constrained. The lower boundary of the mass gap can be as low as  $\sim{}40$ M$_\odot$ or as large as $\sim{}65$ M$_\odot$, depending on details of the stellar evolution model (e.g. with/without rotation, \cite{mapelli2019_2}) and of the pair-instability SN model.

Finally, in the current paper we assume a simplified evolution of the merger rate with redshift (uniform in comoving volume and source-frame time). The evolution with redshift is important (especially for ET) not only for the merger rate but also for the properties of merging systems, because of the influence of the cosmic star formation rate and of the metallicity evolution on BBHs (e.g. \citealt{mapelli2017,2016MNRAS.463L..31L}).  All these caveats must be kept in mind when interpreting the results of our study. Thus, in a follow-up study we will include different metallicities, a self-consistent redshift evolution model, and we will consider a larger parameter space.

Previous papers already addressed the issue of dynamical formation versus isolated formation of BBHs within a model-selection approach (e.g. \citealt{stevenson2015,stevenson2017b,vitale2017,zevin2017}). Most previous studies adopt just simple prescriptions for the dynamical evolution and do not consider a full set of dynamical simulations. \cite{zevin2017}  did compare results of a  population-synthesis sample and a set of dynamical simulations. However, their approach is significantly different from ours as we make use of direct N-body simulations of young star clusters, while they consider hybrid Monte Carlo simulations of globular clusters.
Globular clusters are massive ($>10^4$ M$_\odot$) old star clusters (most of them formed around 12 Gyr ago). They are site of intense dynamical processes: binary hardening and exchanges in globular clusters are very effective (see e.g. \citealt{wang2015,rodriguez2015,askar2017}), but nowadays the stellar mass still locked up in globular clusters is a small fraction of the total stellar mass ($<1$ \%, \citealt{harris2013}). In contrast, young star clusters are smaller systems (the systems we consider here have mass $\sim{}10^{3-4}$ M$_\odot$) and are mostly short-lived ($<1$ Gyr), but they form continuously through the cosmic history and are  expected to host the bulk of massive star formation %
\citep{lada2003,weidner2006,weidner2010}. Thus, the importance of dynamics in a single young star cluster is lower than in a single globular cluster, but the cumulative contribution of young star clusters to the dynamical formation of BBHs is a key factor (e.g. \citealt{dicarlo2019,kumamoto2019}).

From a more technical point of view, globular clusters are spherically symmetric relaxed systems. Hence, they can be simulated with a fast Monte Carlo approach \citep{henon1971,joshi2000}. In contrast, young star clusters are asymmetric and irregular systems, still on their way to relaxation \citep{portegieszwart2010}. Hence, we need more computationally expensive direct N-body simulations to model them realistically. Thus, our approach and the one followed by \cite{zevin2017} are complementary both scientifically and numerically. To understand the dynamical formation of BBHs, we need to model both the globular cluster and the young star cluster environment. The final goal is to have a model-selection tool to distinguish isolated binary formation from dynamical formation, able to account for the many different flavours of dynamical formation (globular clusters, young star clusters, galactic nuclei and hierarchical triples). While we are still far from this goal, our work provides a new crucial piece of information in this direction.

\section{Summary}

The formation channels of BBHs are still an open question. Here, we use a Bayesian model-selection framework and apply it to the isolated binary scenario versus the dynamical scenario of BBH formation in young star clusters. Young star cluster dynamics might be extremely important for BBHs, because the vast majority of massive stars (which are progenitors of BHs) form in young star clusters and OB associations (see e.g. \citealt{lada2003,portegieszwart2010}). However, only few studies focus on BBH formation in young star clusters, because this is a computational challenge \citep{ziosi2014,mapelli2016,banerjee2017,banerjee2018,kumamoto2019,dicarlo2019}. Here, we consider the largest sample of merging BBHs produced in a set of N-body simulations of young star clusters \citep{dicarlo2019}. For the isolated binaries, we take a sample of $>3\times{}10^4$ merging BBHs simulated with the population synthesis code {\sc mobse} \citep{giacobbo2018a,giacobbo2018b}.  {\sc mobse} includes state-of-the-art models for stellar winds and BH formation. The same population-synthesis algorithm is used also in the N-body simulations, ensuring a fair comparison of the two scenarios (see Section \ref{sec_2}).

We analyzed the two scenarios with a Bayesian hierarchical modeling approach capable of estimating which models best fit a given set of GW observations (see Section \ref{sec_3}). We looked at two different cases where we assumed that the underlying astrophysics is either described by a single model, or by a combination of the two models weighted by a mixing fraction parameter $\mathlcal{f}$. In both analyses, the statistical framework was %
applied on the combination of the mass parameters and the effective spin. %

In terms of GW observations, we used both mock data and LVC observations during O1 and O2 \citep{abbottO2}.  Our results with mock observations showed that the distributions of $M_{t}$ and $q$ already present strong features that can be used to differentiate between the two models. In fact, with $~500$ observations with aLIGO and aVirgo we could be able to restrict the values of the mixing fraction to an interval smaller than $0.5$. With the inclusion of the effective spin parameter in the analysis, this interval becomes even smaller, with values close to $0.15$. 

Finally, this work is the first one that used the latest LVC data to perform Bayesian model selection approach in order to discriminate between BBHs formed via isolated or dynamical binaries %
Our results showed that the current set of observations is not able to put a strong constraint on the mixing fraction of the the two models. A pure isolated (dynamical) scenario in which all BBH progenitors have metallicity $Z=0.002$, as described by our simulations, is barely consistent (still consistent) with LVC data, because of the presence of massive BBHs such as GW170729. We stress that progenitor metallicity and dynamics have a somewhat degenerate effect on the maximum mass of merging BBHs: we expect the pure isolated scenario will still be consistent with O1+O2 data if we include more metal-poor progenitors (down to $Z\sim{}0.0002$). Thus, in a follow-up study, we will apply our methodology to a range of metallicities for both the dynamical and isolated scenarios.

Finally, given our estimations obtained with mock observations, we expect that after about a hundred detections (optimistic scenario for O3, middle panel of Figure \ref{fig:mcmc_gen_obs}), we should already be able to constrain the values of the mixing fraction in an interval smaller than $0.5$.

\section*{Acknowledgments}
MM and YB  acknowledges financial support by the European Research Council for the ERC Consolidator grant DEMOBLACK, under contract no. 770017. YB would like to acknowledge networking support by the COST Action GWverse CA16104.
EB and VB are supported by NSF Grant No. PHY-1841464, NSF Grant No. AST-1841358, NSF-XSEDE Grant No. PHY-090003, and NASA ATP Grant No. 17-ATP17-0225. 
This work has received funding from the European Union’s Horizon 2020 research and innovation programme under the Marie Skłodowska-Curie grant agreement No. 690904. 
Computational work was performed on the University of Birmingham's BlueBEAR cluster and at the Maryland Advanced Research Computing Center (MARCC).

\newpage{}

\bibliography{Bouffanais_ApJ}

\end{document}